\documentstyle[12pt,openbib,psfig,amssymb,amsmath,graphicx]{article}
\begin{document}

\title{\bf Rotational parameters of strange stars in comparison with neutron stars}

\author{Manjari Bagchi$^{1,~*}$}

\maketitle

{$^1$ Tata Institute of Fundamental Research, Colaba, Mumbai 400005, India \\
 $^*$ manjari@tifr.res.in}

\begin{abstract}

I study stellar structures $i.e.$ the mass, the radius, the moment of inertia and the oblateness parameter at different spin frequencies for strange stars and neutron stars in a comparative manner. I also calculate the values of the radii of the marginally stable orbits and Keplerian orbital frequencies. By equating kHz QPO frequencies to Keplerian orbital frequencies, I find corresponding orbital radii. Knowledge about these parameters might be useful in further modeling of the observed features from LMXBs with advanced and improved future techniques for observations and data analysis.

\end{abstract}

\vskip .5cm

\noindent Keywords : {dense matter; equation of state; X-rays: binaries; stars: neutron; stars: rotation} \\

\noindent PACS : {26.60.Kp; 97.10.Nf; 97.10.Pg; 97.60.Jd; 97.80.Jp; 97.10.Kc}

\newpage

\section{Introduction}
\label{sec:intro}

Presently there are a number of efforts to constrain the dense matter Equations of State (EsoS) through astronomical observations of compact stars. The usual approach is to determine the mass and the radius of the stars with the help of various observational features like gravitational redshifts (z) from spectral lines, cooling characteristics, kHz quasi-periodic oscillations (QPO) etc (Lattimer \& Prakash 2007, Li $et~al.$ 1999, \"Ozel 2006, 2008, Zhang $et~al.$ 2007). But these methods are not foolproof, $e.g.$ the value of z used in \"Ozel's (2006) analysis of EXO 0748$-$676 can not be reproduced as mentioned by Klahn $et~al.$ (2006). Moreover, to constrain EsoS from QPO observations, one needs to believe in a particular model of QPO which is again a subject of debate. Another alternative method might be the measurement of the moment of inertia from the faster component of the double pulsar system PSR J0737-3039 (Lattimer \& Schutz 2005, Bagchi $et~al.$ 2008).  Some high mass stars like PSR J1903+0327, EXO 0748$-$676 $etc.$ prefer stiff EoS and some other stars like 4U 1728-34 (Li $et~al.$ 1999), EXO 1745-248 (\"Ozel 2008), prefer soft EoS. This fact hints to the possibility of existence of both neutron stars and strange stars. But even then, I need some constrains as there are a number of EsoS for neutron stars and also for strange stars. Until then, it is interesting to compare the stellar properties for different EsoS. For sufficiently fast spinning stars, stellar structures depend upon the spin frequency ($\nu_{spin}$). So the study of stellar structures for rotating stars will help in better understanding of the characteristics of fast spinning compact stars like LMXBs and millisecond pulsars. That is why here I study stellar structures with rotations in section \ref{sec:stell_struct}. In section \ref{sec:rot_param}, I study different rotational parameters for strange stars and neutron stars. Although stellar structures for rotating neutron stars or strange stars have been already studied by a number of groups (some of which I discuss in section \ref{sec:rot_param}), systematic studies of all relevant stellar parameters as well as disk parameters were lacking. That is why here I report  the variation of a number of different stellar parameters as well as disk parameters for different values of star's spin frequency and mass in section \ref{sec:stell_struct} and section \ref{sec:rot_param}. Also I use one EoS for strange stars and another EoS for neutron stars whereas in the earlier works people discussed either only neutron star rotations or only strange star rotations, there was no comparison between the properties of rotating neutron stars and rotating strange stars. In addition I compare my results obtained by using a pseudo-Newtonian potential with full general relativistic calculations by other people like Haensel \& Zdunik (1989) and Lattimer \& Prakash (2004) and the close matching found implies the correctness of my approach and the validity of the pseudo-Newtonian potential. In section \ref{sec:application} I discuss a possible application of the knowledge of the rotational parameters in modeling kHz QPOs. I end with a discussion in section \ref{sec:conclusion}.

\section{Stellar structures with rotation}
\label{sec:stell_struct}
\begin{figure}
\centerline{\psfig{figure=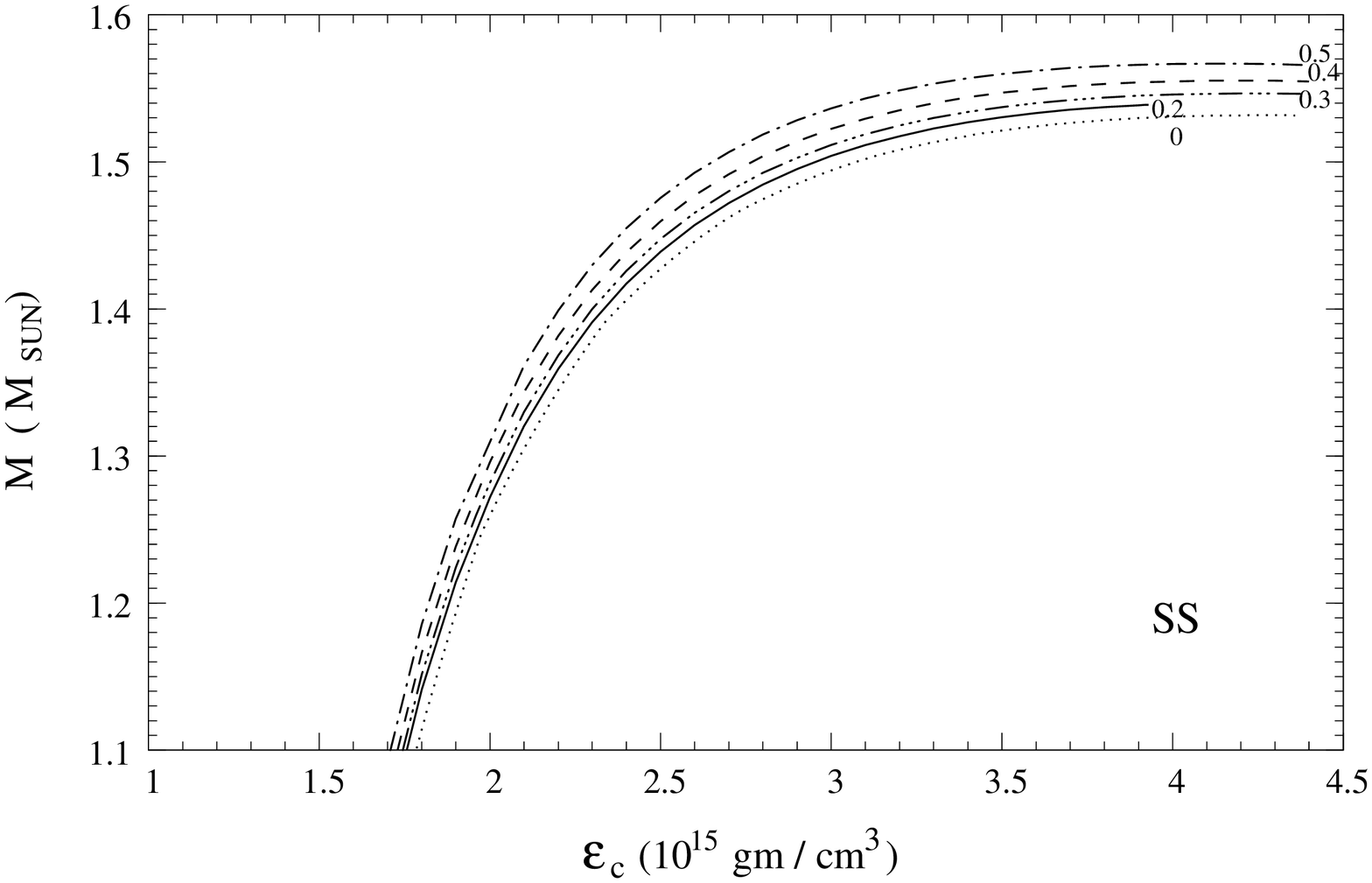,width=8cm}}
\centerline{\psfig{figure=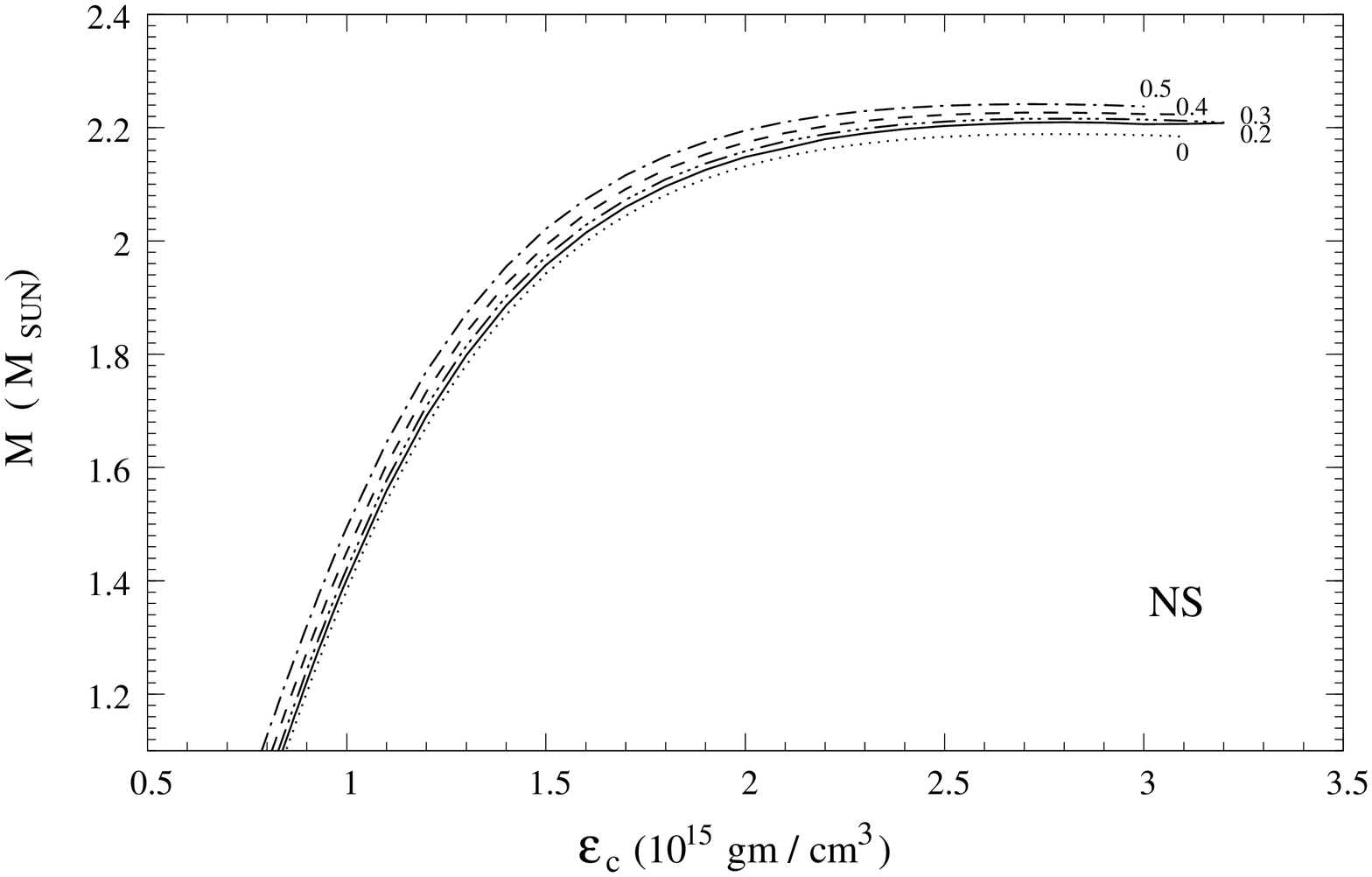,width=8cm}}
\caption{Variation of the mass with the central density for strange stars (upper panel) and neutron stars (lower panel). The parameter is the value of $\Omega$ in units of $ 10^{4}~{\rm sec}^{-1}$. The EsoS used are EoS A for strange stars and EoS APR for neutron stars. \label{fig:em}} 
\end{figure}

\begin{figure}
\centerline{\psfig{figure=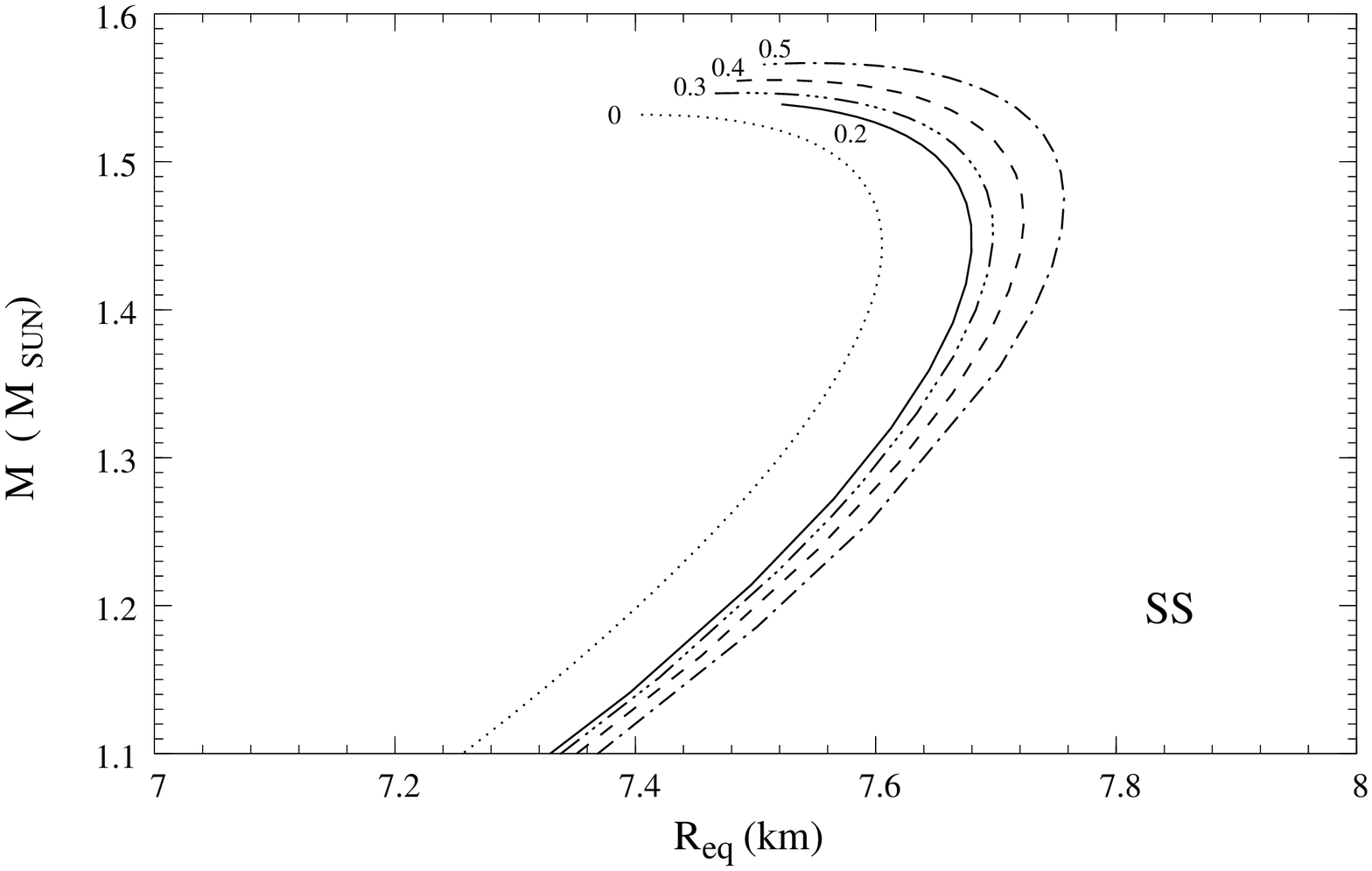,width=8cm}}
\centerline{\psfig{figure=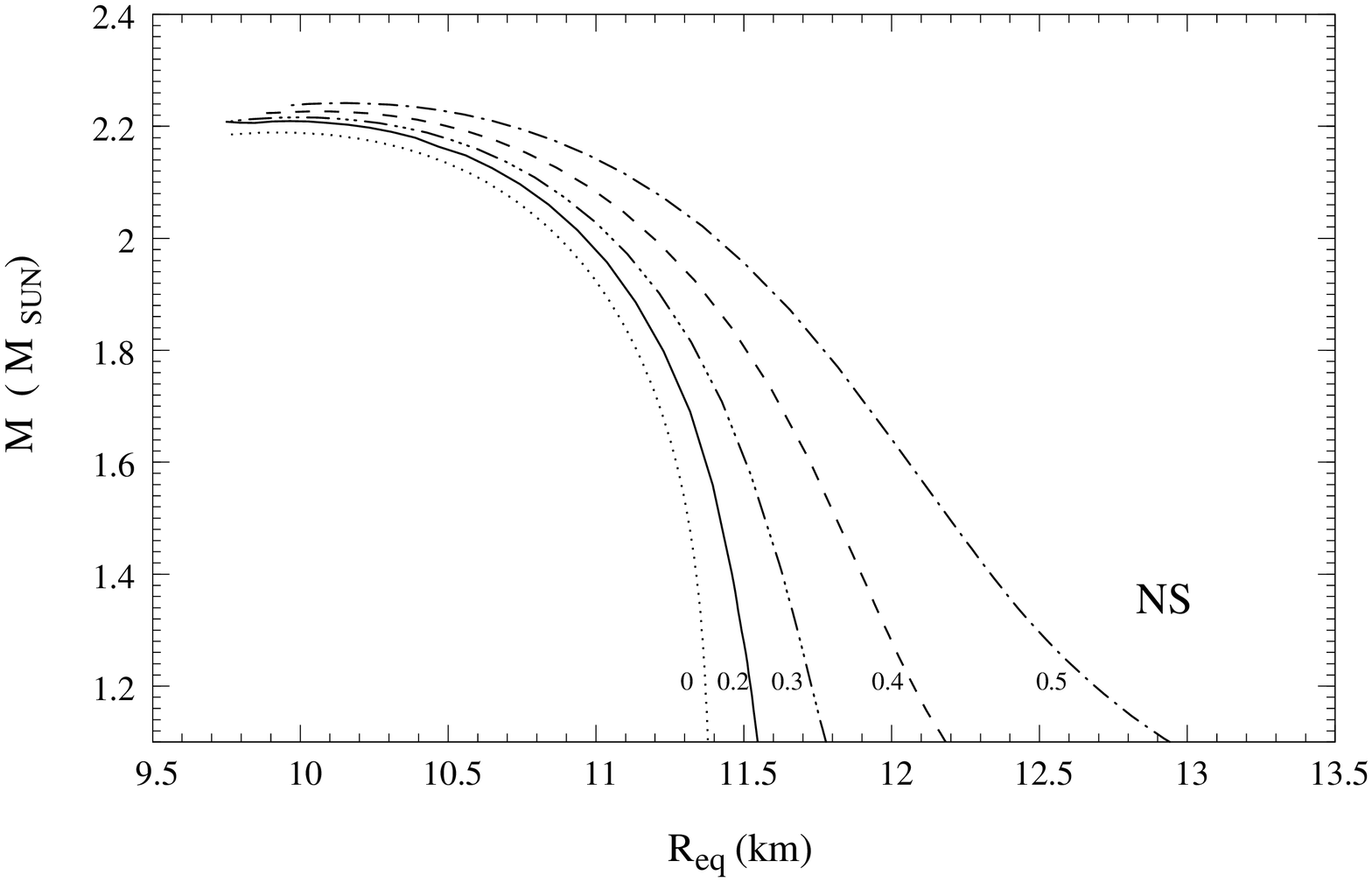,width=8cm}}
\caption{Variation of the mass with the radius for strange stars (upper panel) and neutron stars (lower panel). The parameter is the value of $\Omega$ in units of $ 10^{4}~{\rm sec}^{-1}$. The EsoS used are EoS A for strange stars and EoS APR for neutron stars.  \label{fig:rm}}
\end{figure}

\begin{figure}
\centerline{\psfig{figure=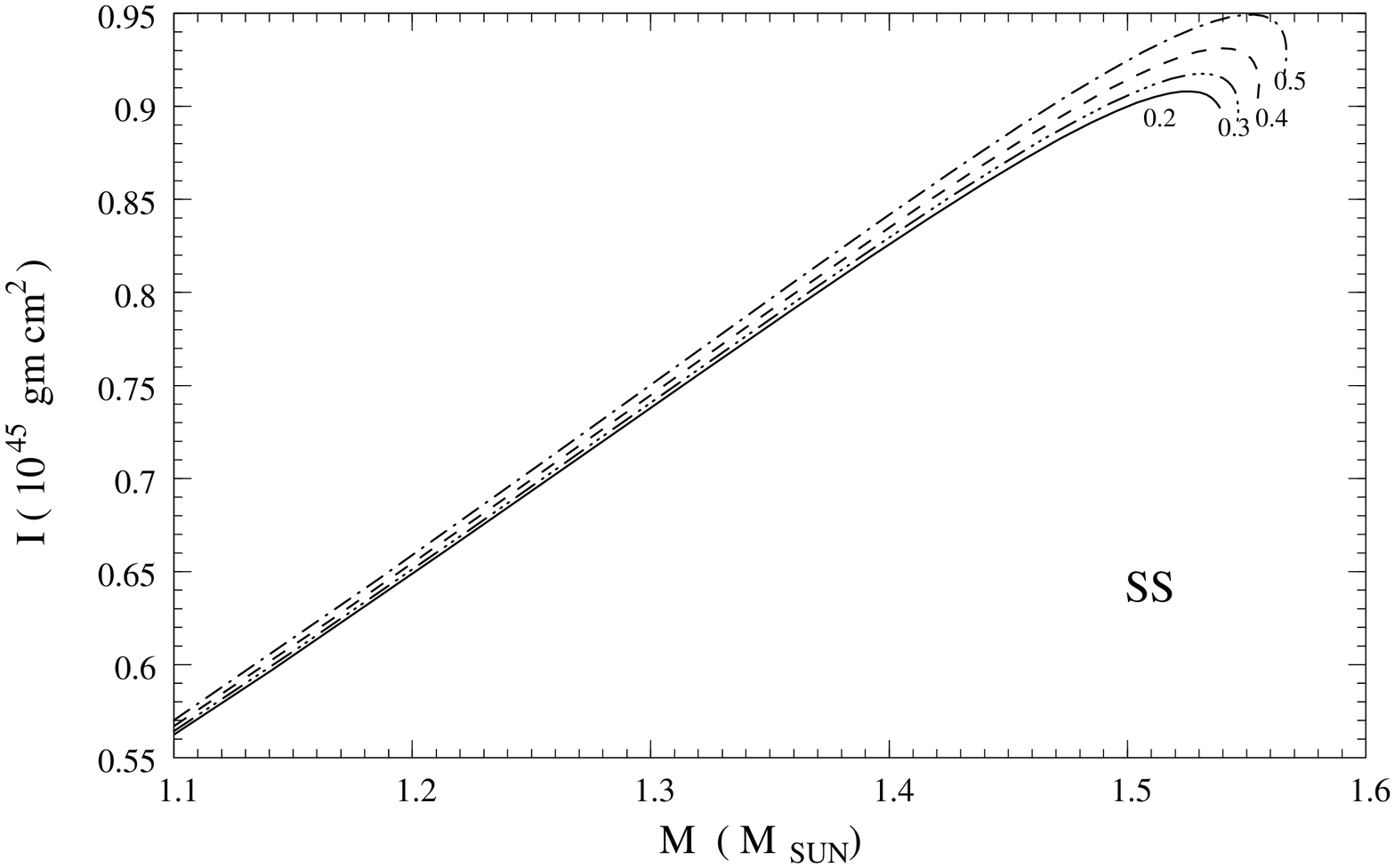,width=8cm}}
\centerline{\psfig{figure=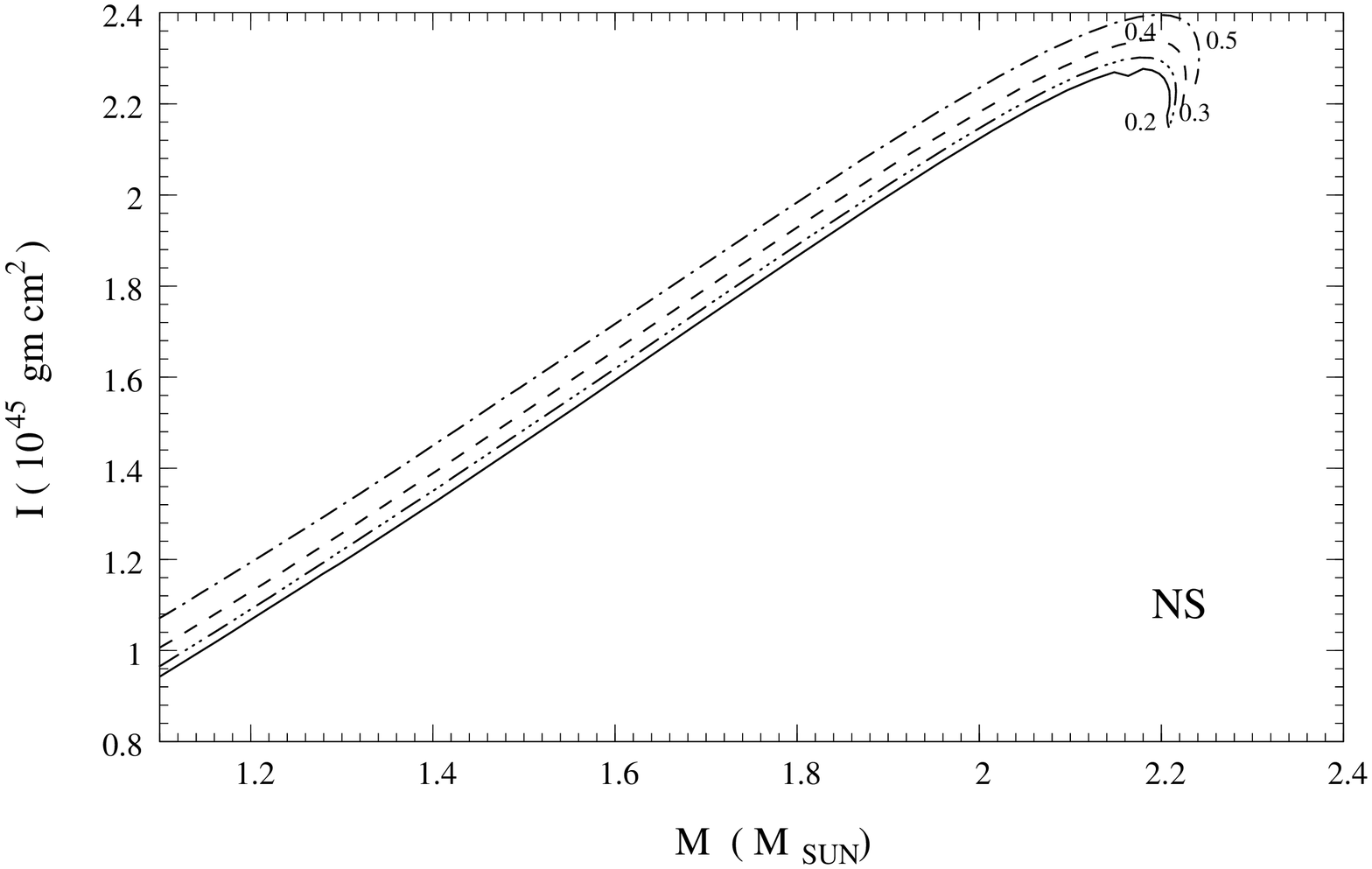,width=8cm}}
\caption{Variation of the moment of inertia with the mass for strange stars (upper panel) and neutron stars (lower panel). The parameter is the value of $\Omega$ in units of $ 10^{4}~{\rm sec}^{-1}$. The EsoS used are EoS A for strange stars and EoS APR for neutron stars. \label{fig:mi}}
\end{figure}

\begin{figure}
\centerline{\psfig{figure=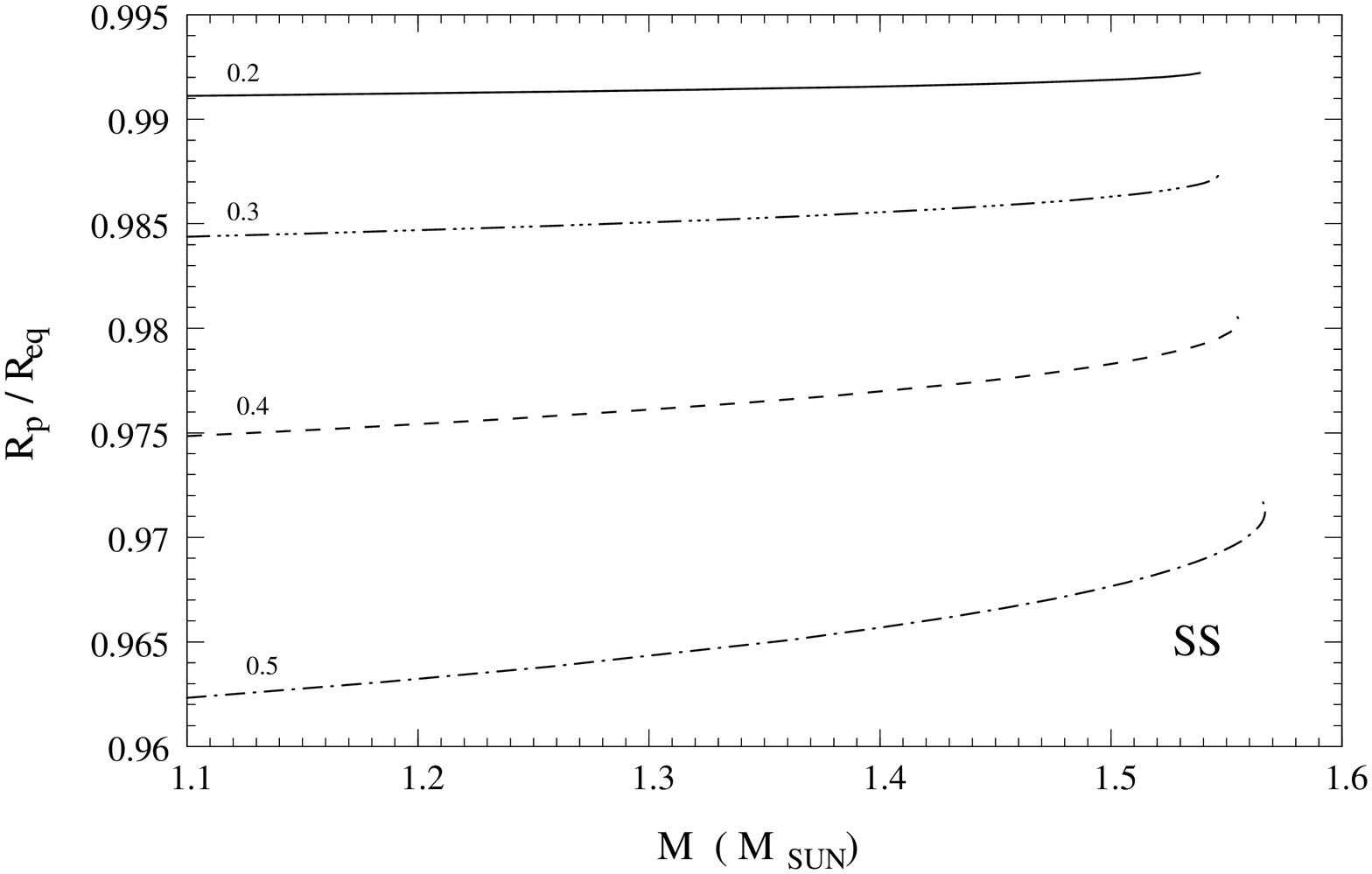,width=8cm}}
\centerline{\psfig{figure=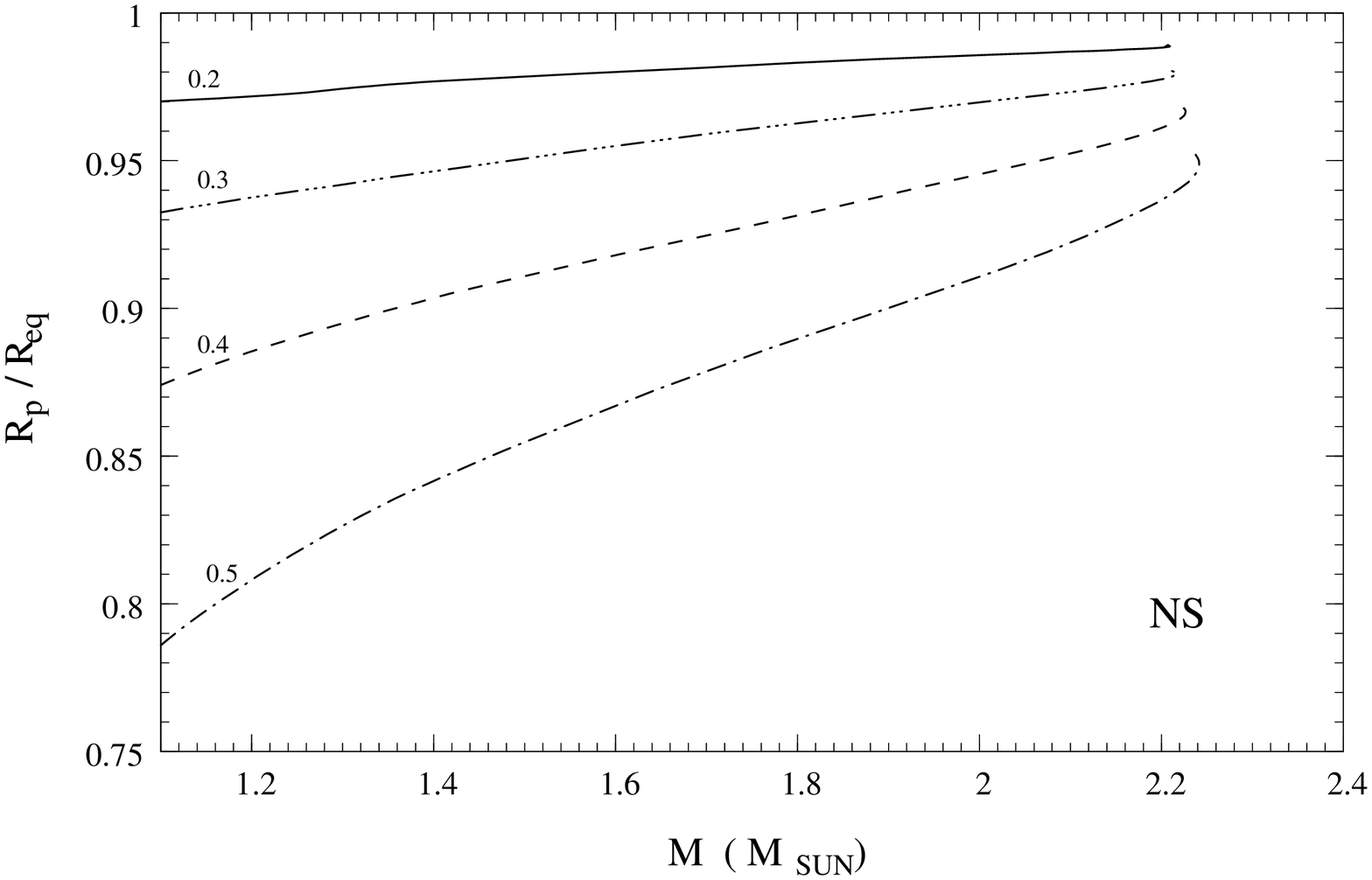,width=8cm}}
\caption{Variation of the oblateness parameter $i.e.$ the ratio of the polar radius to the equatorial radius with the central density for strange stars (upper panel) and neutron stars (lower panel). The parameter is the value of $\Omega$ in units of $ 10^{4}~{\rm sec}^{-1}$. The EsoS used are EoS A for strange stars and EoS APR for neutron stars. \label{fig:mob}}
 \label{fig:mass_all_eos}
\end{figure}

\begin{figure}
\centerline{\psfig{figure=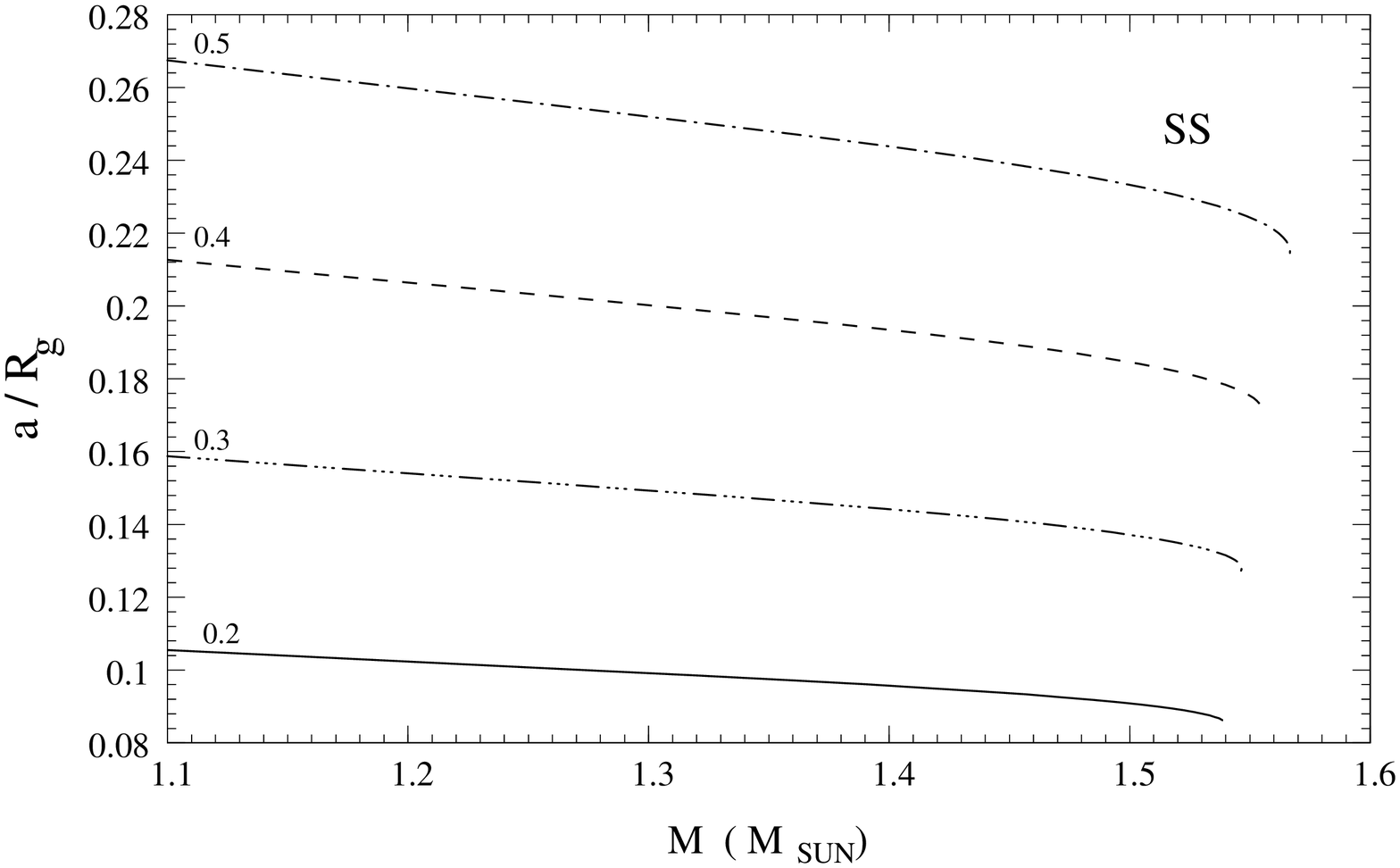,width=8cm}}
\centerline{\psfig{figure=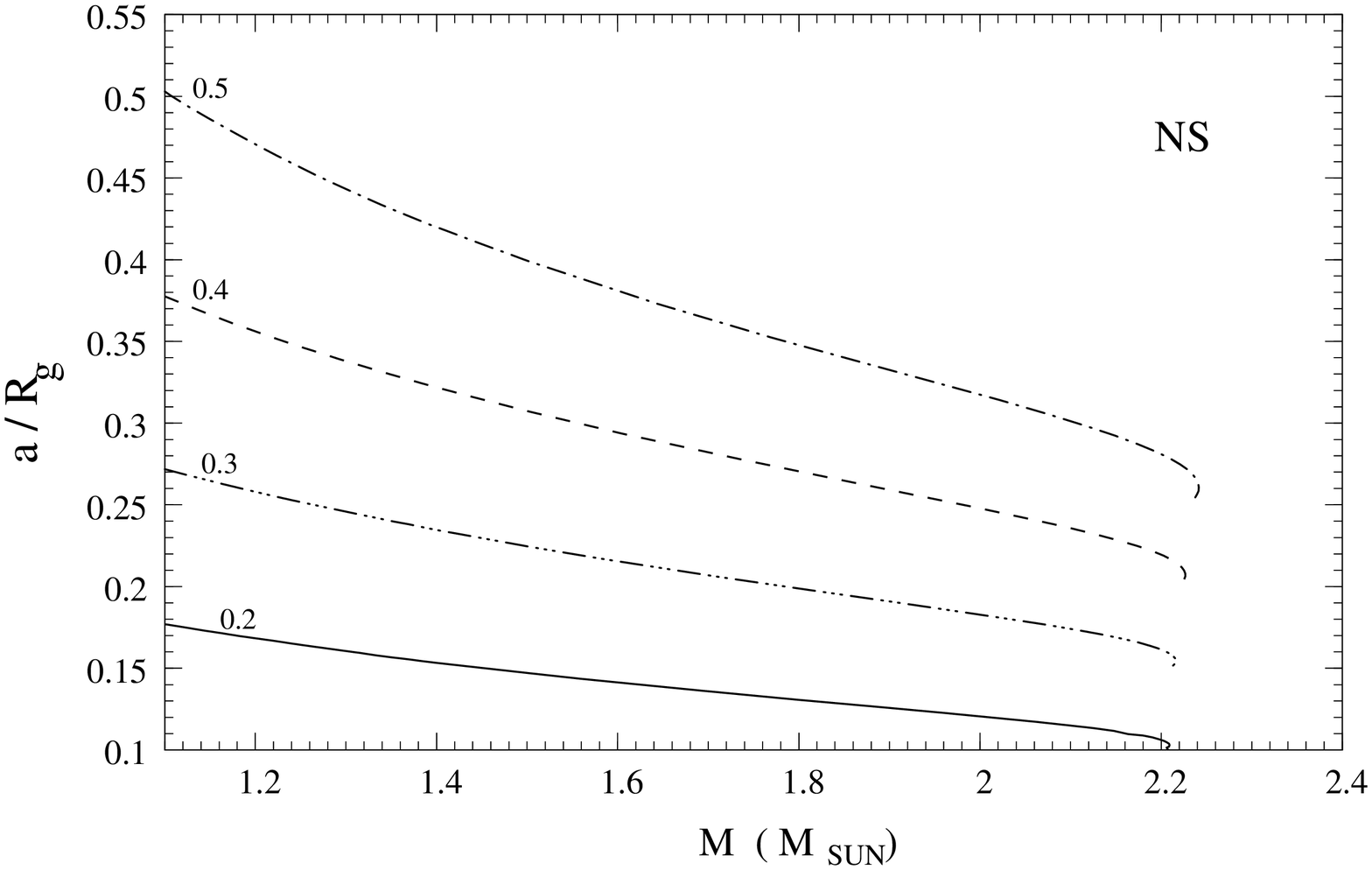,width=8cm}}
\caption{Variation of $a/R_g$ with the mass for strange stars (upper panel) and neutron stars (lower panel). The parameter is the value of $\Omega$ in units of $ 10^{4}~{\rm sec}^{-1}$. The EsoS used are EoS A for strange stars and EoS APR for neutron stars.  \label{fig:ma}}
\end{figure}

I use two sample EsoS of the dense matter among the numerous EsoS available in literature, one for the strange quark matter (EoS A or SSA, Bagchi $et~al.$, 2006) and the other for the nuclear matter (EoS APR Akmal, Pandharipande \& Ravenhall 1998). To find stellar structures with rotations, I use the RNS code\footnote{http://www.gravity.phys.uwm.edu/rns/}. Following Komatsu, Eriguchi \& Hachisu (1989), this code constructs the compact star models by solving stationary, axisymmetric, uniformly rotating perfect fluid solutions of the Einstein field equations with tabulated EsoS (supplied by the users). 

The fastest rotating compact star known till date is probably XTE J1739-285 (Kaaret $et~al.$ 2007) having $\nu_{spin}=1122$ Hz, although the measurement has not been confirmed later. The second fastest one is J1748-2446ad (Hessels $et~al.$ 2006) with $\nu_{spin}= 716$ Hz.  In this work, I choose the angular frequency ($\Omega~=~2 \pi \nu_{spin})$ as 2000, 3000, 4000 and 5000 $sec^{-1}$ (which correspond to $\nu_{spin}$ as $\sim$ 318 Hz, 477 Hz, 637 Hz and 796 Hz respectively) . All of the fast rotating compact stars except XTE J1739-285 have $\nu_{spin}$ in that range. I have also computed non-rotating, spherically symmetric stellar structures by solving TOV equations which are sufficient for slow objects like EXO 0748$-$676 ($\nu_{spin}=45$ Hz). Throughout this work, I take the stellar mass ($M$) to be always greater than 1.1 $M_{\odot}$ as observations usually hint the stellar mass to be greater than that value.

In Fig \ref{fig:em} I plot the mass against the central density ($\epsilon_c$) both for strange stars and neutron stars. For a fixed value of $\epsilon_c$, $M$ increases little bit with the increase of $\Omega$. For all of the values of $\Omega$, $M$ first increases with the increase of $\epsilon_c$ ($\frac{\partial M}{\partial \epsilon_c}>0$) and then after a certain value of $M$ ($M_{max}$) starts to decrease ($\frac{\partial M}{\partial \epsilon_c}<0$). The stars are unstable when $\frac{\partial M}{\partial \epsilon_c}<0$.  This instability appears around $\epsilon_c= 4.1 \times 10^{15}~\rm{gm~cm^{-3}}$ for strange stars and around $\epsilon_c= 2.8 \times 10^{15}~\rm{gm~cm^{-3}}$ for neutron stars; these values does not change more than $5\%$ with the change of $\Omega$ in the chosen range. For any $\Omega$ in the chosen range, I get $M=1.1~M_{\odot}$ at $\epsilon_c \sim 1.7 \times 10^{15}~\rm{gm~cm^{-3}}$ for strange stars and at $\epsilon_c \sim 0.80 \times 10^{15}~\rm{gm~cm^{-3}}$ for neutron stars.

Fig \ref{fig:rm} shows the mass-radius plots. With the increase of $\Omega$, for both strange stars and neutron stars, the maximum mass ($M_{max}$) increases and for any fixed mass, the radius also increases due to the larger value of the centrifugal force. Note that here ``radius" means the equatorial radius $R_{eq}$ which is always greater than the polar radius $R_{p}$. For a fixed $\Omega$, the maximum mass for a neutron star is greater than that of a strange star. For fixed values of $\Omega$ and $M$, $R_{eq}$ is larger for a neutron star than that of a strange star. The compactness factor ($M/R$) of strange stars is larger than that of neutron stars and the variation of $M$ with $R_{eq}$ follows an approximate $R_{eq}^{3}$ law for strange stars in contrast to neutron star's approximate $R_{eq}^{-3}$ variation. 

Fig \ref{fig:mi} shows the variation of the moment of inertia ($I$) with the mass. For any fixed mass, the moment of inertia increases with the increase of $\Omega$ both for strange stars and neutron stars. For fixed values of $M$ and $\Omega$, a neutron star possess much higher value of $I$ than a strange star because of its larger value of $R_{eq}$.

In Fig \ref{fig:mob}, I plot the oblateness parameter $i.e.$ the ratio of the polar radius to the equatorial radius ($R_{p}/R_{eq}$) with the mass. It is clear that for a fixed mass, this ratio decreases with the increase of $\Omega$ both for strange stars and neutron stars $i.e.$ the star becomes more and more oblate due to the larger values of the centrifugal force. Moreover, for a fixed value of $\Omega$, this ratio decreases with the decrease of the mass as there the centrifugal force becomes increasingly more dominant over the gravitational force. Note that the variation of $R_{p}/R_{eq}$ with $M$ is steeper for neutron stars than that for strange stars, but for both of them, the steepness increases with the increase of $\Omega$.

In Fig \ref{fig:ma}, I plot $a/R_g$ with $M$ where $R_g~=~GM/c^2$ and $a~=~I\Omega/Mc$. As expected from their expressions, the plot shows that for any fixed mass, $a/R_g$ increases with the increase of $\Omega$ as expected and for a fixed $\Omega$, $a/R_g$ decreases with the increase of the mass. For the same value of $M$ and $\Omega$, a neutron star has larger value of $a/R_g$ than that of a strange star because of its larger value of $I$. $a/R_g$ is an important parameter of the compact stars as it can be identified as the specific angular momentum of the star and its value determines many other properties of the star.

For any other EoS, the value of $M_{max}$ and corresponding radius will change depending upon the stiffness of that EoS. But the general trend of the M-R curve will remain the same $i.e.$ $M \propto R_{eq}^3$ for strange stars and $M \propto R_{eq}^{-3}$ for neutron stars. The nature of $M-\epsilon_c$ curve will also remain the same.

\section{Rotational parameters}
\label{sec:rot_param}

With the output of the RNS code, $i.e.$ $M,~R_{eq}$ and $I$, I calculate some rotational parameters for strange stars and neutron stars. First I calculate the radius of the marginally stable orbit which is defined as (Bardeen $et~al.$, 1972) :

\begin{equation} 
r_{ms}~=~ R_g \{3+Z_2 \mp [(3-Z_1)(3+Z_1+2Z_2)]^{1/2}\}\label{eq:rms}
\end{equation}
where
\begin{equation} 
Z_1~=~ 1+[1-(a/R_g)^2]^{1/3}[(1+a/R_g)^{1/3}+(1-a/R_g)^{1/3}]\label{eq:z1}
\end{equation}
and 
\begin{equation} 
Z_2~=~ [3(a/R_g)^2+Z_1^2]^{1/3}\label{eq:z2}
\end{equation}

The ``-" sign in the expression of $r_{ms}$ implies the co-rotating motion and the ``+" sign implies the counter-rotating motion which I call as $r_{ms,~co}$ and $r_{ms,~counter}$ respectively. As the values of $a/R_g$ are always very small, both $Z1$ and $Z2$ have their values $\sim$ 3.

The Keplerian frequency of a particle orbiting around the star at a radial distance $r$ can be expressed as

\begin{equation}
\nu_{k}(r)~=~\frac{1}{2 \pi}\left[\frac{F_{m}(r)}{r}\right]^{1/2}\label{eq:nuk}
\end{equation}
where $F_{m}(r)$ is the force per unit mass. As an example, I take $F_{m}(r)$ as derived from a pseudo-Newtonian potential by Mukhopadhyay \& Misra (2003)

\begin{equation}
F_{m}(r)~=~\frac{R_g c^2}{r^2}\left[1-\left(\frac{r_{ms}}{r}\right)+\left(\frac{r_{ms}}{r}\right)^2 \right]\label{eq:fm}
\end{equation}

\begin{figure}
\centerline{\psfig{figure=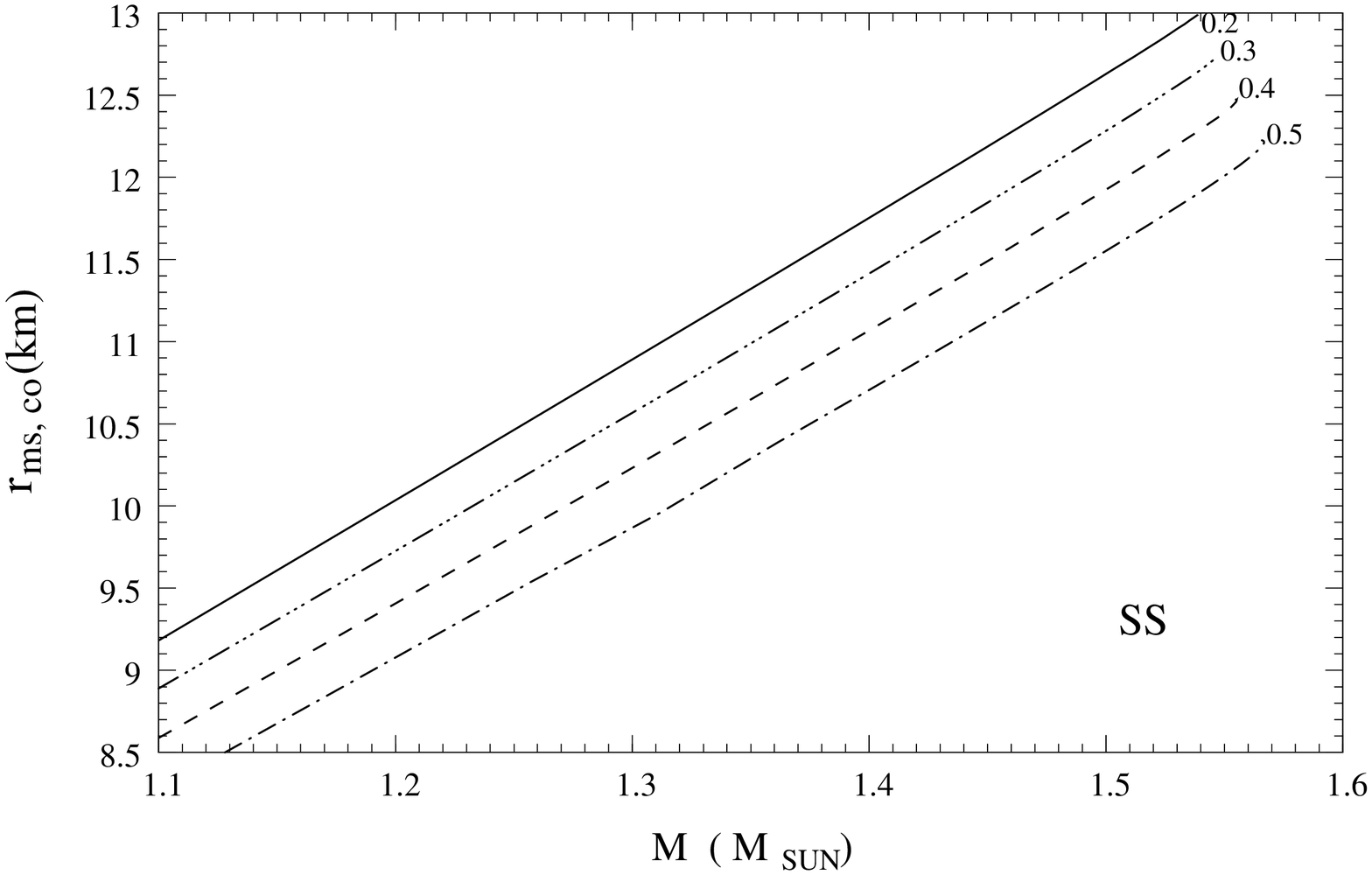,width=8cm}}
\centerline{\psfig{figure=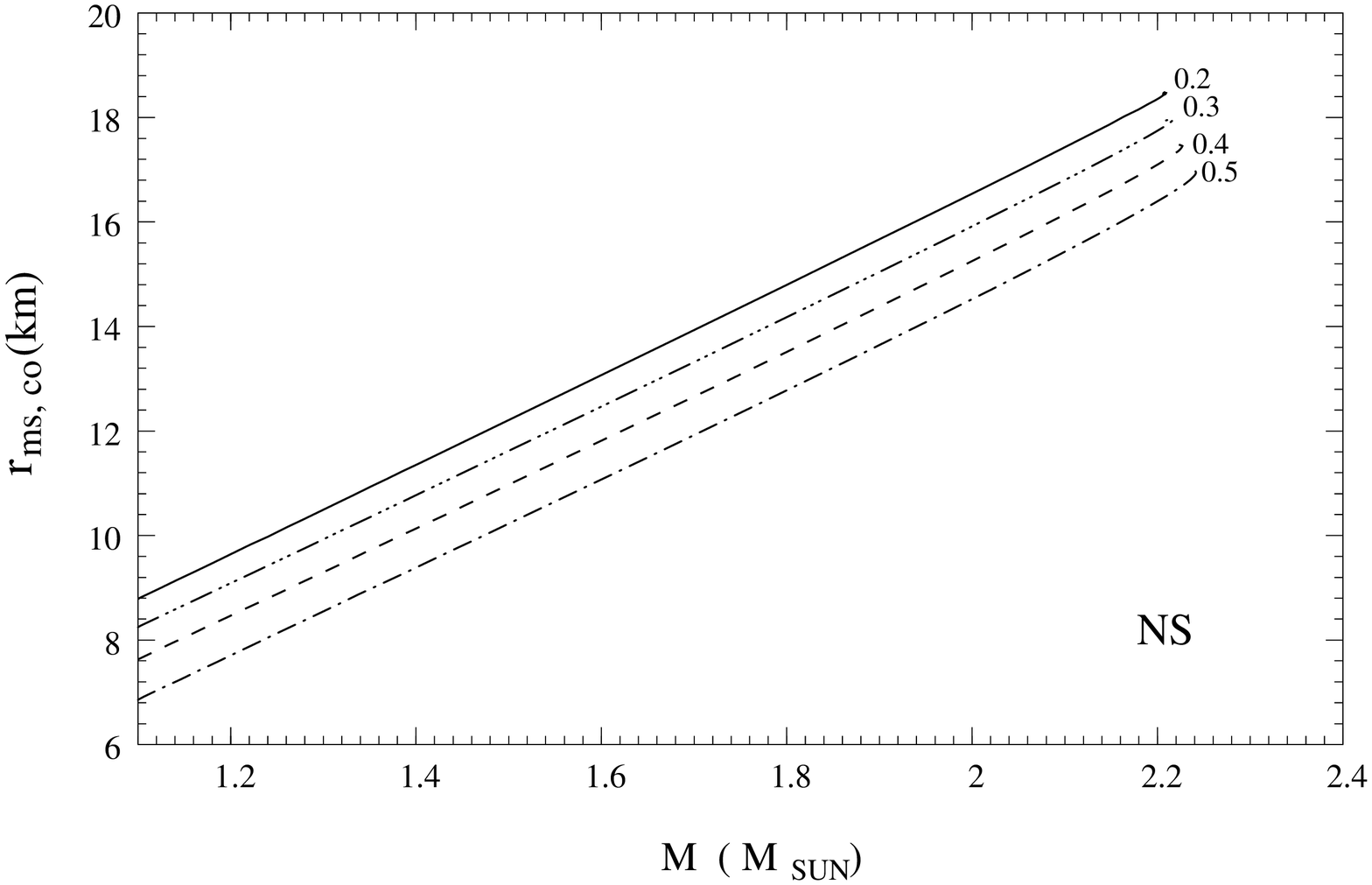,width=8cm}}
\caption{Variation of $r_{ms,~co}$ with the mass for strange stars (upper panel) and neutron stars (lower panel). The values of $\Omega$ in units of $ 10^{4}~{\rm sec}^{-1}$ are shown. The EsoS used are EoS A for strange stars and EoS APR for neutron stars. \label{fig:m_rmsco}}
\end{figure}

\begin{figure}
\centerline{\psfig{figure=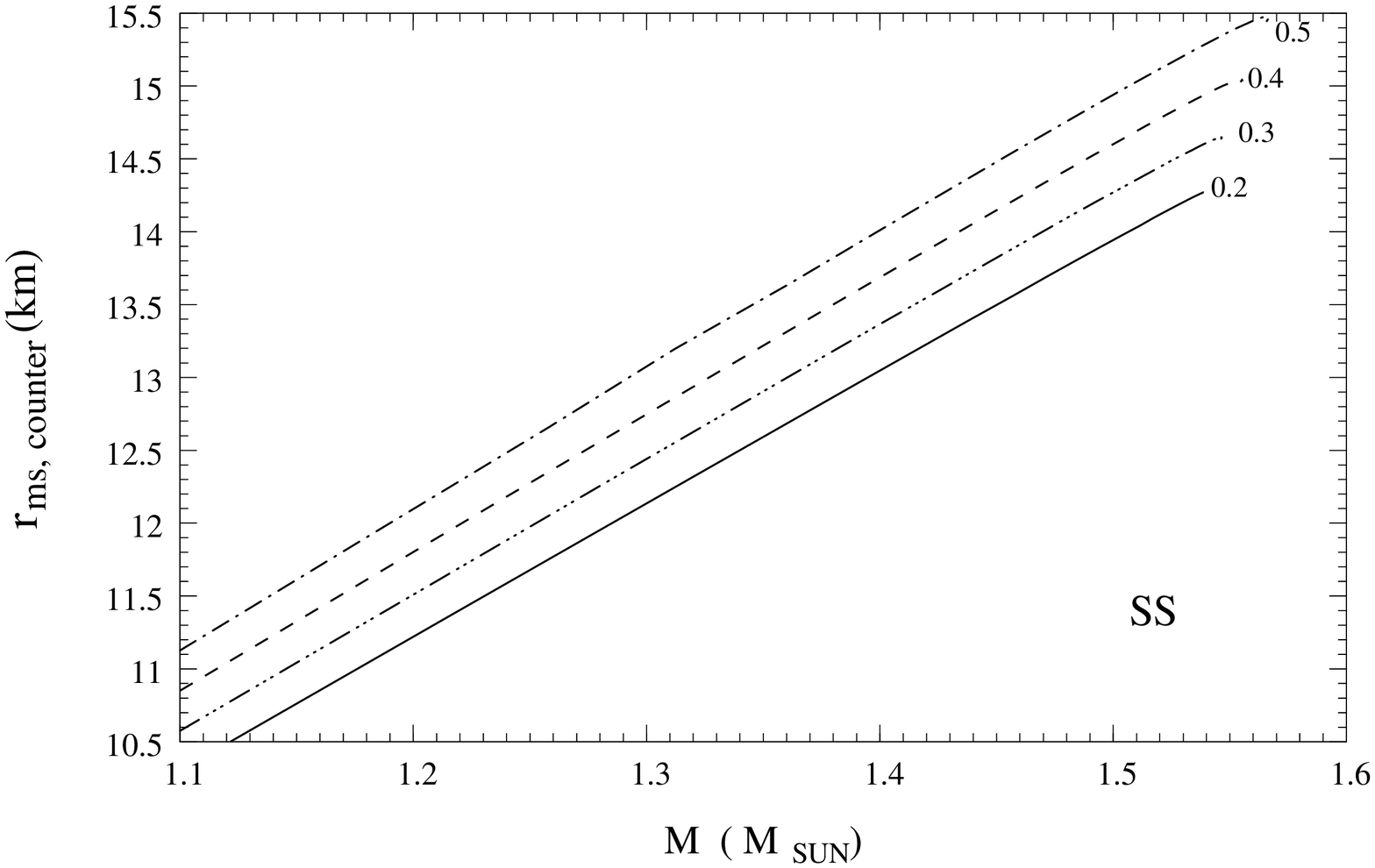,width=8cm}}
\centerline{\psfig{figure=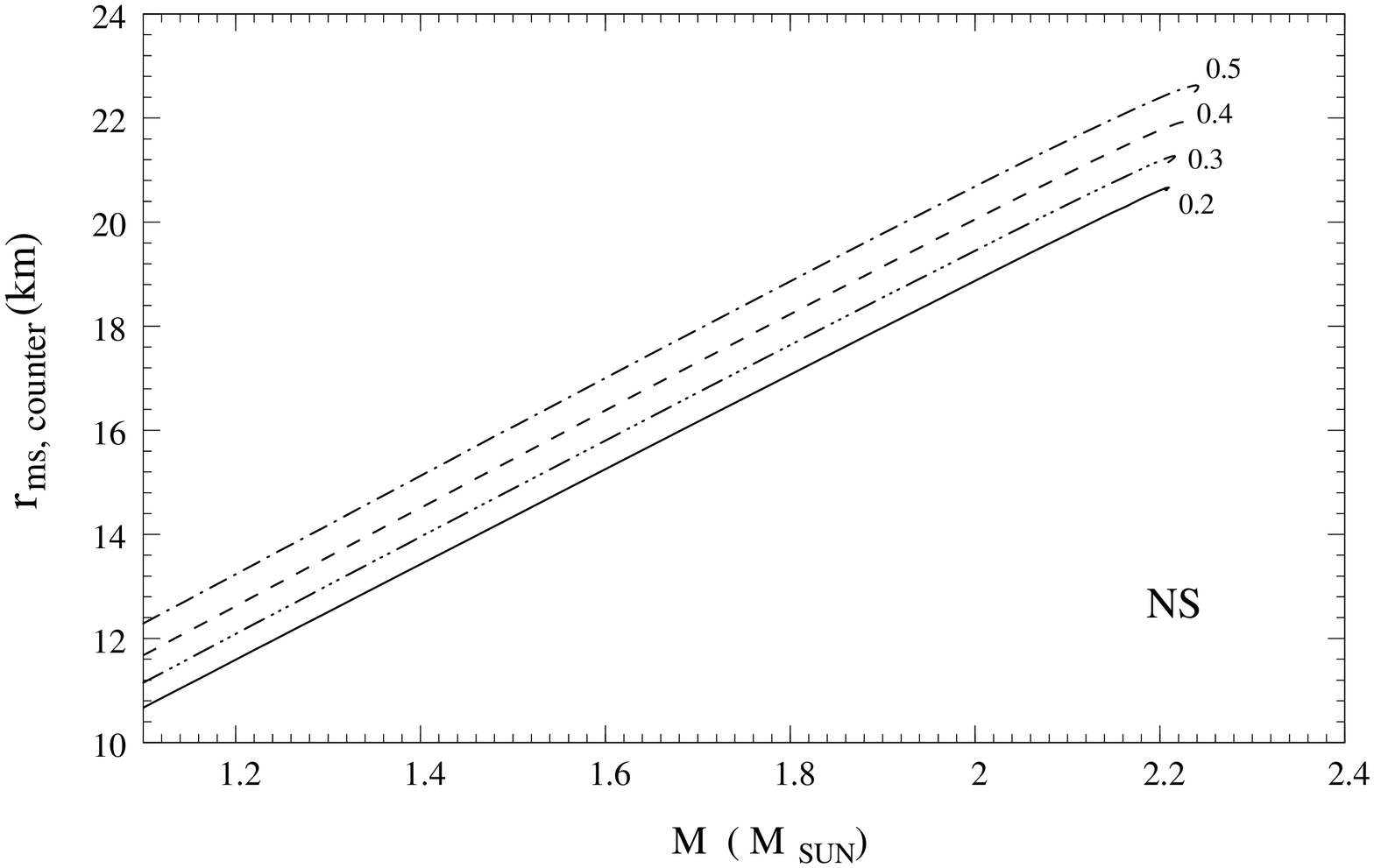,width=8cm}}
\caption{Variation of $r_{ms,~counter}$ with the mass for strange stars (upper panel) and neutron stars (lower panel). The parameter is the value of $\Omega$ in units of $ 10^{4}~{\rm sec}^{-1}$. The EsoS used are EoS A for strange stars and EoS APR for neutron stars. \label{fig:m_rmscounter}}
\end{figure}

In Figs \ref{fig:m_rmsco} and \ref{fig:m_rmscounter}, I plot $r_{ms,~co}$ and $r_{ms,~counter}$ respectively with the mass. For any EoS, $r_{ms,~co}$ is always smaller than $r_{ms,~counter}$ for any fixed values of $\Omega$ and $M$. For a fixed $\Omega$, both $r_{ms,~co}$ and $r_{ms,~counter}$ increases linearly with the increase of $M$. But for a fixed $M$, $r_{ms,~co}$ decreases with the increase of $\Omega$ and $r_{ms,~counter}$ increases with the increase of $\Omega$. For strange stars, both $r_{ms,~co}$ and $r_{ms,~counter}$ are always greater than $R_{eq}$. For neutron stars, $r_{ms,~co}$ and $r_{ms,~counter}$ are smaller than $R_{eq}$ for low  masses and they become greater than $R_{eq}$ for higher masses; $r_{ms,~co}$ becomes equal to $R_{eq}$ around $1.5-1.7 M_{\odot}$ and $r_{ms,~counter}$ becomes equal to $R_{eq}$ around $1.15-1.2M_{\odot}$.

All these happen because of the nature of the terms in the expression of $r_{ms}$ (Eqn.\ref{eq:rms}). As both $R_g \{3+Z_2\}$ and $R_g \{[(3-Z_1)(3+Z_1+2Z_2)]^{1/2}\}$ are positive quantities and the second one is smaller than the first one, their sum ($r_{ms,~counter}$) must be greater than their difference ($r_{ms,~co}$). As for a fixed value of $\Omega$, both $R_g \{3+Z_2\}$ and $R_g \{[(3-Z_1)(3+Z_1+2Z_2)]^{1/2}\}$ increases with the increase of $M$ with the first term having much steeper slope, both $r_{ms,~co}$ and $r_{ms,~counter}$ increase with increase of $M$. For a fixed mass, $R_g \{3+Z_2\}$ remains almost constant but $R_g \{[(3-Z_1)(3+Z_1+2Z_2)]^{1/2}\}$ increases with the increase of $\Omega$, so their sum ($r_{ms,~counter}$) increases with the increase of $\Omega$ and the difference ($r_{ms,~co}$) decreases with the increase of $\Omega$.

For strange stars (EoS A), $R_g \{3+Z_2\} \gg R_{eq}$ for any value of $M$. So after addition or subtraction of a comparatively small term ($R_g \{[(3-Z_1)(3+Z_1+2Z_2)]^{1/2}\}$) with it, the expression ($r_{ms,~counter}$ or $r_{ms,~co}$) remain always greater than $R_{eq}$

For neutron stars (EoS APR), $R_g \{3+Z_2\}< R_{eq}$ at lower values of $M$ where the values for $R_{eq}$ are sufficiently larger, but $R_g \{3+Z_2\} > R_{eq}$ at larger values of $M$. Here $R_g \{[(3-Z_1)(3+Z_1+2Z_2)]^{1/2}\}$ is very small in comparison to both $R_g \{3+Z_2\}$ and $R_{eq}$ for all values of $M$. So the addition or subtraction of $R_g \{[(3-Z_1)(3+Z_1+2Z_2)]^{1/2}\}$ with $R_g \{3+Z_2\}$ (to get $r_{ms}$ for counter-rotating or co-rotating motions respectively) does not change the overall trend, only the addition (for counter-rotation) shifts the transition towards lower values of M whereas the subtraction does the reverse thing. On the other hand, for strange stars, $R_g \{3+Z_2\}$ is always much greater than $R_{eq}$ and even after the addition or subtraction of the smaller term $R_g \{[(3-Z_1)(3+Z_1+2Z_2)]^{1/2}\}$, it ($r_{ms}$) remains greater than $R_{eq}$. 

\begin{figure}
\centerline{\psfig{figure=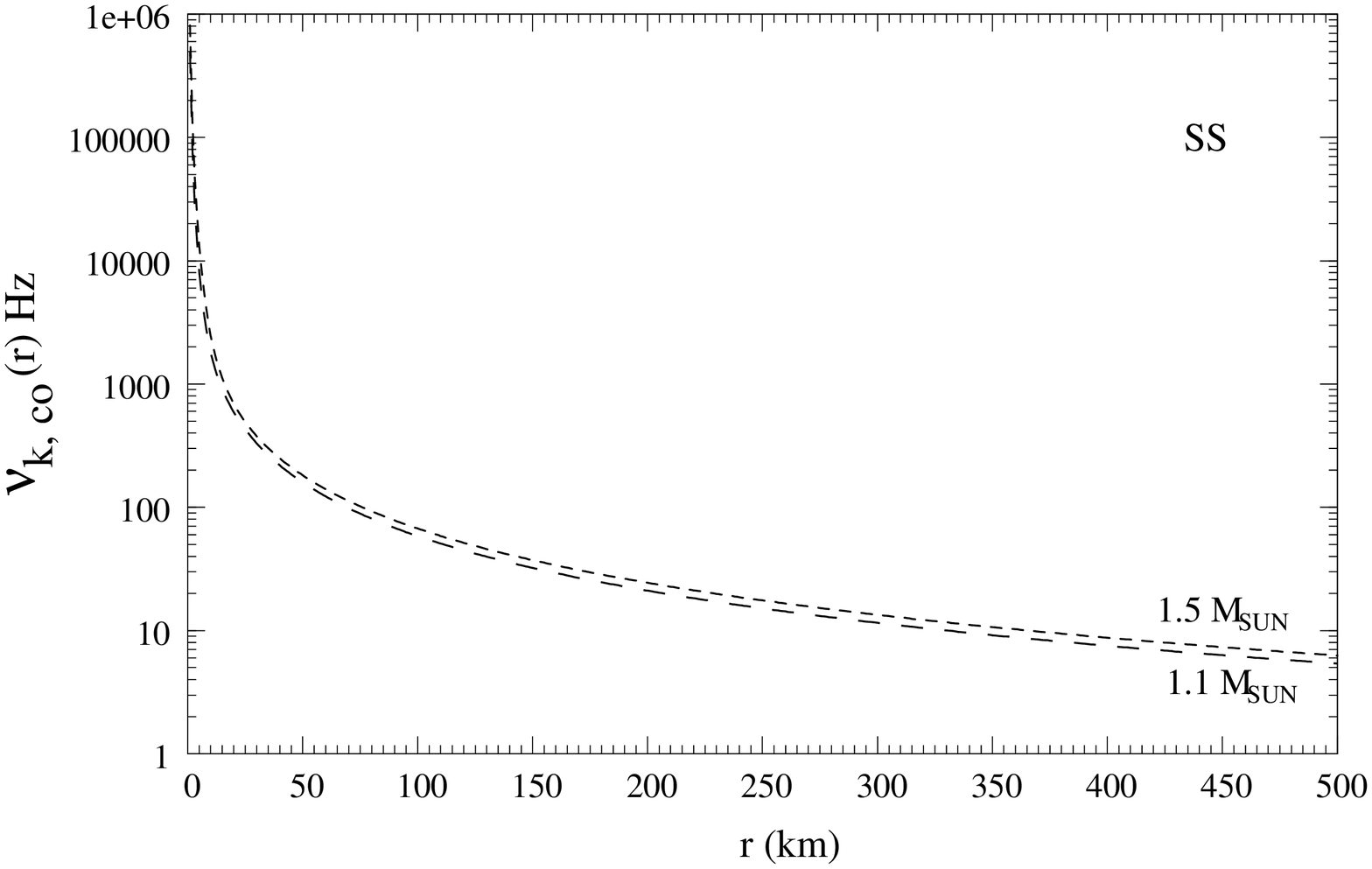,width=8cm}}
\centerline{\psfig{figure=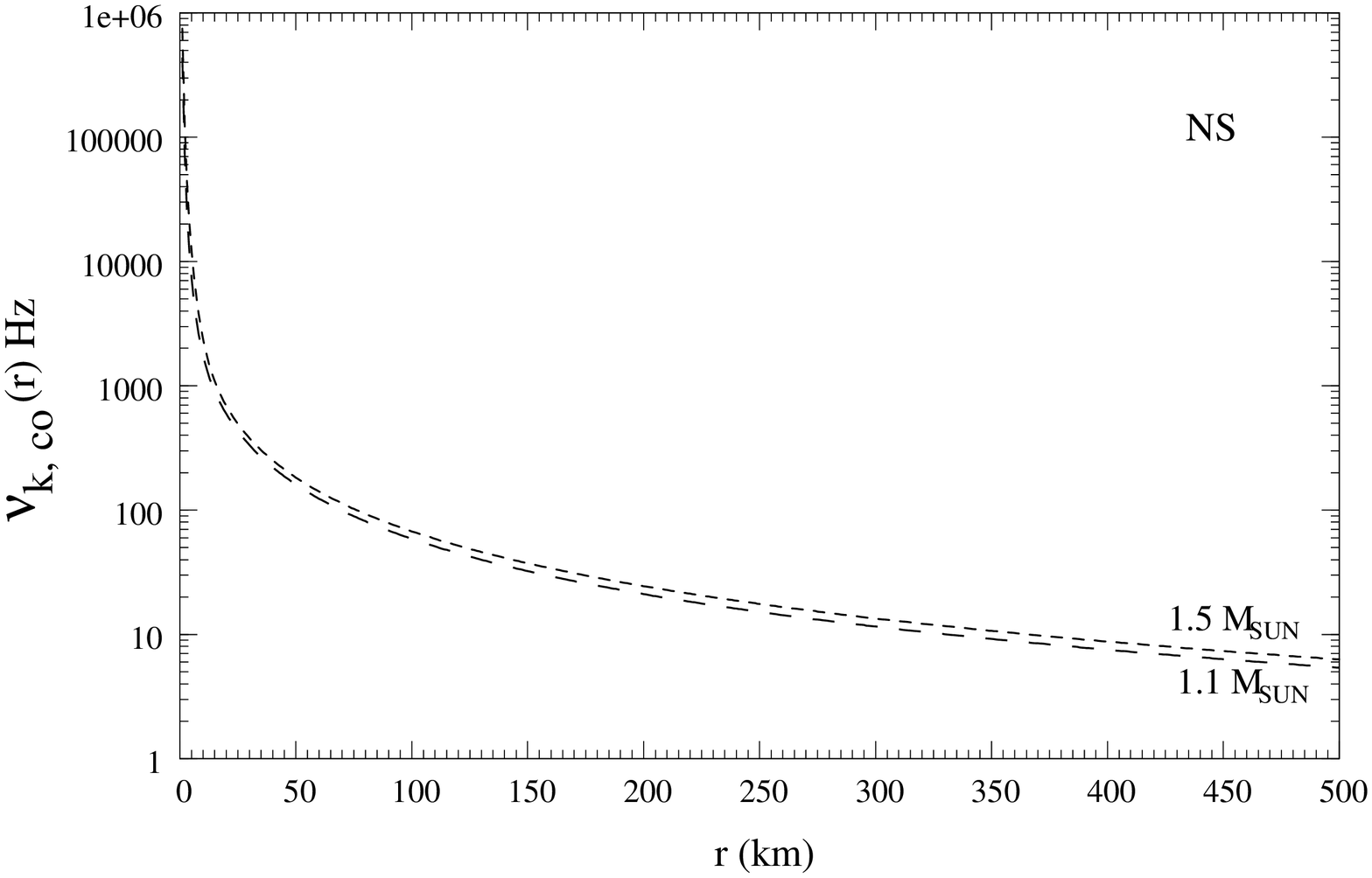,width=8cm}}
\caption{Variation of $\nu_{k}(r)$ with the mass for strange stars (upper panel) and neutron stars (lower panel) for $\Omega~=~0.4 \times 10^{4}~{\rm sec}^{-1}$. Stellar masses are taken as 1.1 $M_{\odot}$ and 1.5 $M_{\odot}$.  The EsoS used are EoS A for strange stars and EoS APR for neutron stars. \label{fig:rnuco}}
\end{figure}

In Fig \ref{fig:rnuco}, I plot the variation of $\nu_{k}(r)$ as a function of $r$.  No significant difference in $\nu_{k}(r)$ (at any chosen $r$)  between a strange star and a neutron star having the same values of $M$ and $\Omega$ is observed (specially at higher values of $r$). In comparison to the whole range of $\nu_{k}(r)$, the differences between $\nu_{k}(r)$ for the co-rotating and the counter-rotating motions (keeping all of the other parameters fixed) and the variation of  $\nu_{k}(r)$ with $\Omega$ (keeping all of the other parameters fixed) are very small. $ | \left[\nu_{k,~co}(r)-\nu_{k,~counter}(r) \right]/ \left[\nu_{k,~counter}(r) \right] |$ varies around 0.2 to 0.001 for $r=$ 0 to 500 km depending slightly upon the choice of the EoS, $M$ and $\Omega$; $ | \left[\nu_{k,~ 0.4 }(r)-\nu_{k,~0.2}(r) \right] / \left[\nu_{k,~0.2}(r) \right]| $ varies 0.1  to 0.001 for $r=$ 0 to 500 km (where the 3rd parameter in the subscript denotes the value of $\Omega$ in units of $10^{4}~{\rm sec}^{-1}$) depending slightly upon the choice of the EoS, $M$ and the direction of the motion. So as an example, the plot in Fig \ref{fig:rnuco} is only for co-rotating motion with $\Omega~=~0.4 \times 10^{4}~{\rm sec}^{-1}$ and for only two chosen values of $M$. 

But these differences between $\nu_{k, co}(r)$ and $\nu_{k, counter}(r)$  become larger for smaller values of $r$ where $\nu_{k, co}(r) < \nu_{k, counter}(r) $, but at sufficiently higher values of $r$, $\nu_{k, co}(r) \gtrsim \nu_{k, counter}(r) $. This transition occurs at $ r \sim 26-27$ km for both strange stars and neutron stars with $M~=~ 1.5~M_{\odot}$ and $\Omega=0.4 \times 10^4$. Similar trends have been noticed for other values of $M$ and $\Omega$. The value of $r$ at which this transition occurs increases slightly with the increase of $M$ but does not depend significantly with $\Omega$. 

We will now study the variation of $\nu_{k}(R_{eq})$ $i.e.$ $\nu_{k}(r)$ at $r=R_{eq}$ with $M$ taking different values of $\Omega$ and both co-rotating and counter-rotating motions. $\nu_{k, co}(R_{eq})$ is the Keplerian frequency of a particle orbiting the star at the star's equatorial surface and I obtain it by equating the gravitational force with the centrifugal force. 

\begin{figure}
\centerline{\psfig{figure=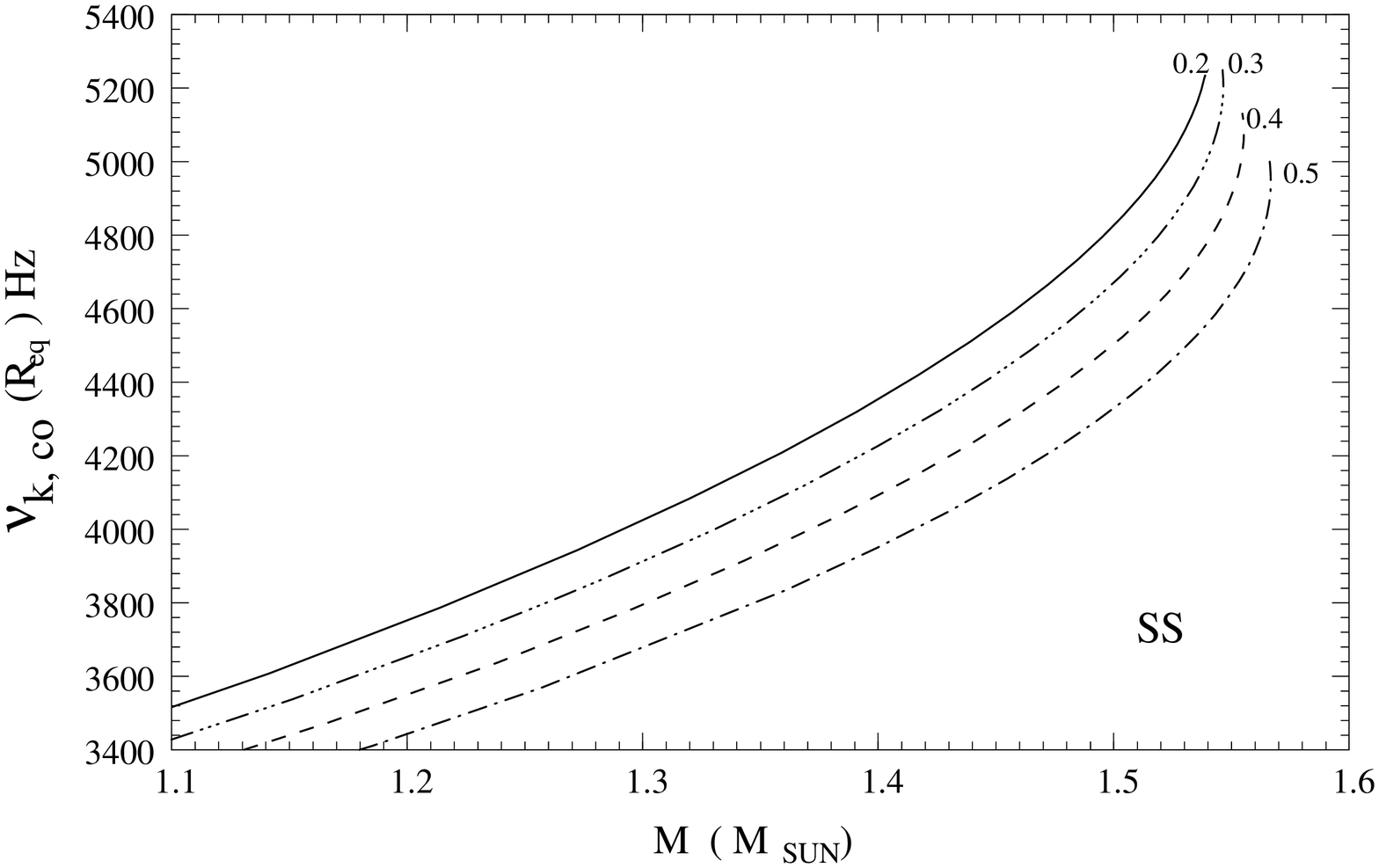,width=8cm}}
\centerline{\psfig{figure=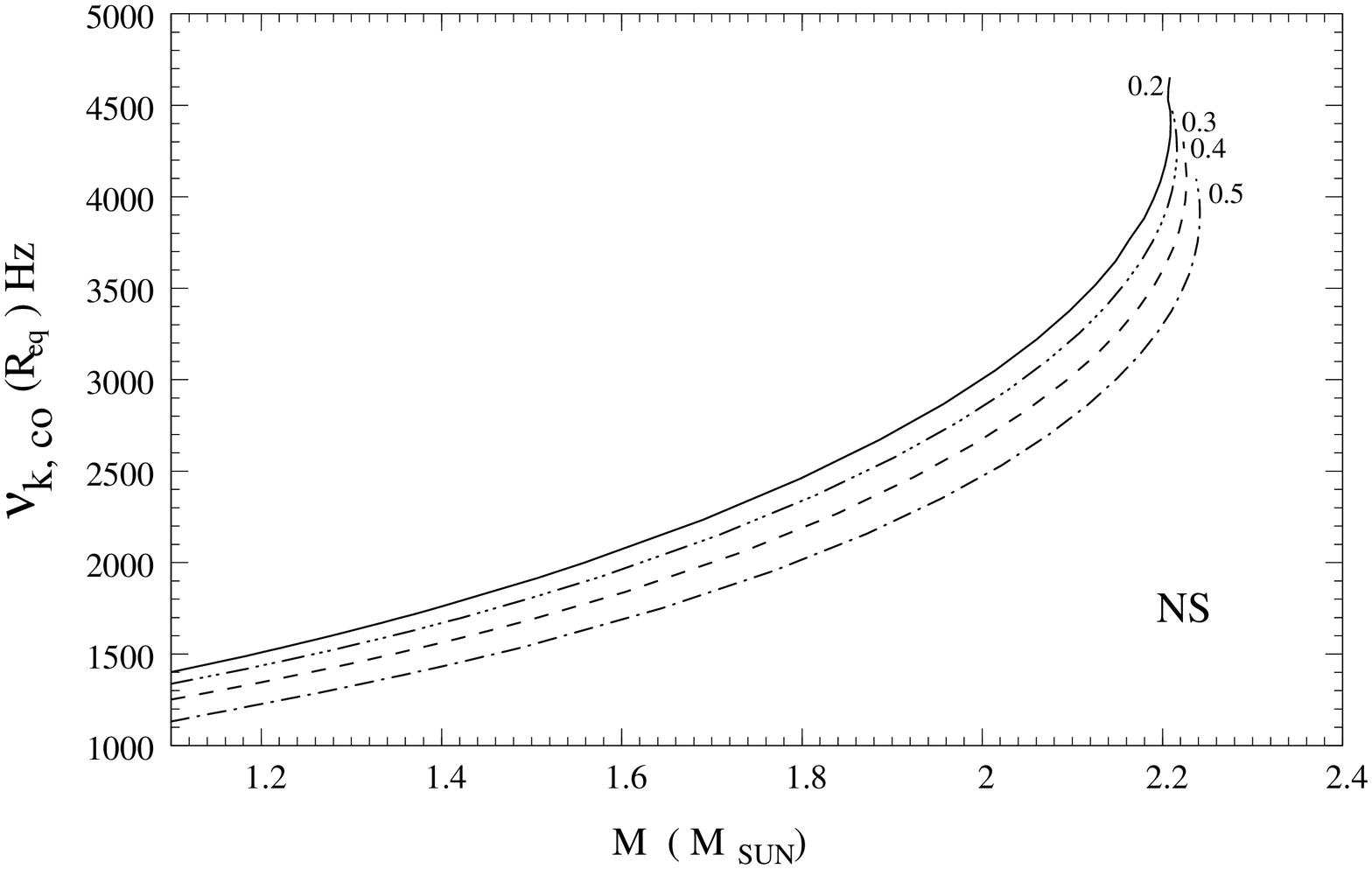,width=8cm}}
\caption{Variation of $\nu_{k,~co}(R_{eq})$ with the mass for strange stars (upper panel) and neutron stars (lower panel). The parameter is the value of $\Omega$ in units of $ 10^{4}~{\rm sec}^{-1}$ corresponding to $\nu_{spin}$ 318 Hz, 477 Hz, 637 Hz and 796 Hz respectively. The EsoS used are EoS A for strange stars and EoS APR for neutron stars. \label{fig:mnukreqco}}
 \end{figure}

\begin{figure}
\centerline{\psfig{figure=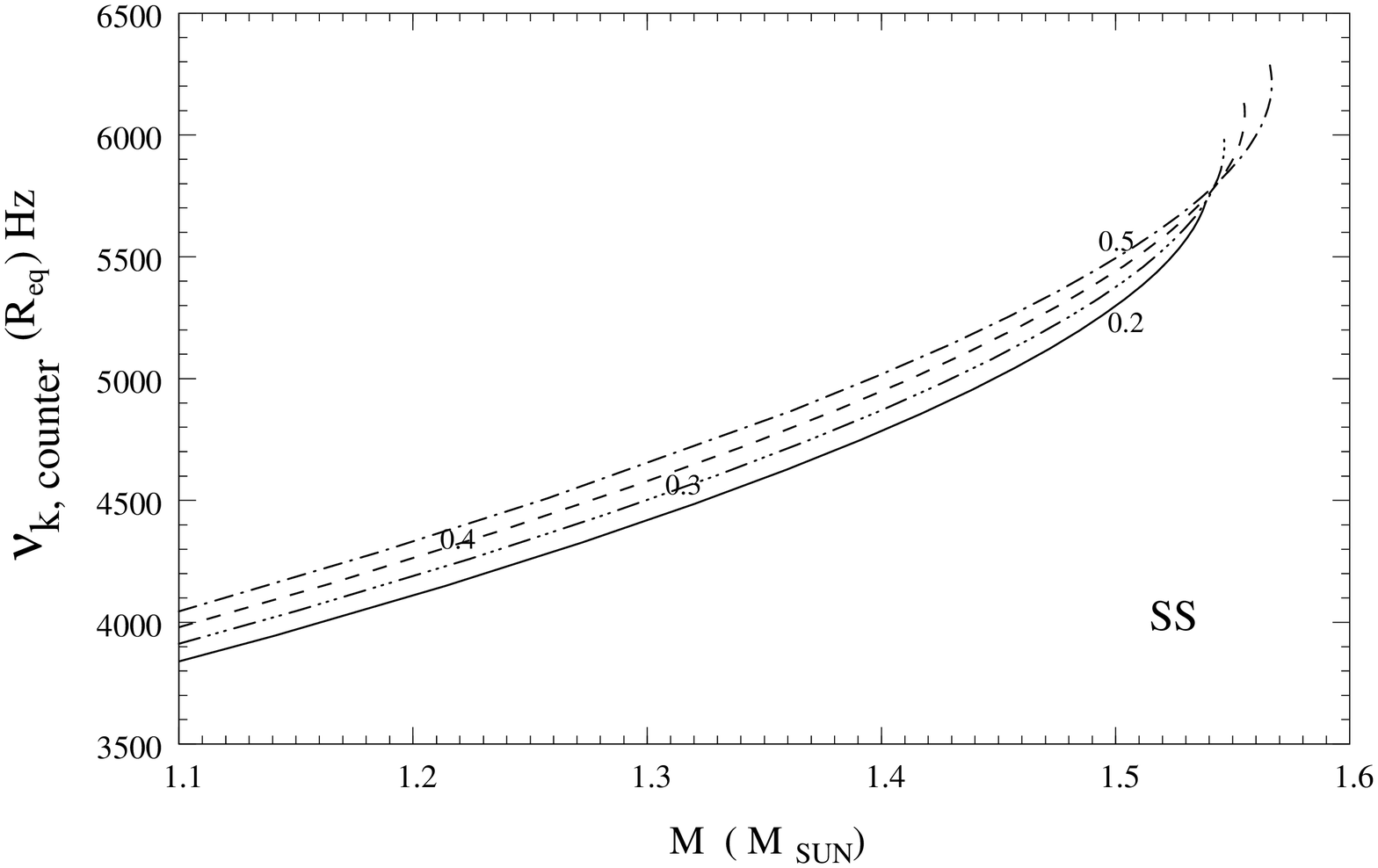,width=8cm}}
\centerline{\psfig{figure=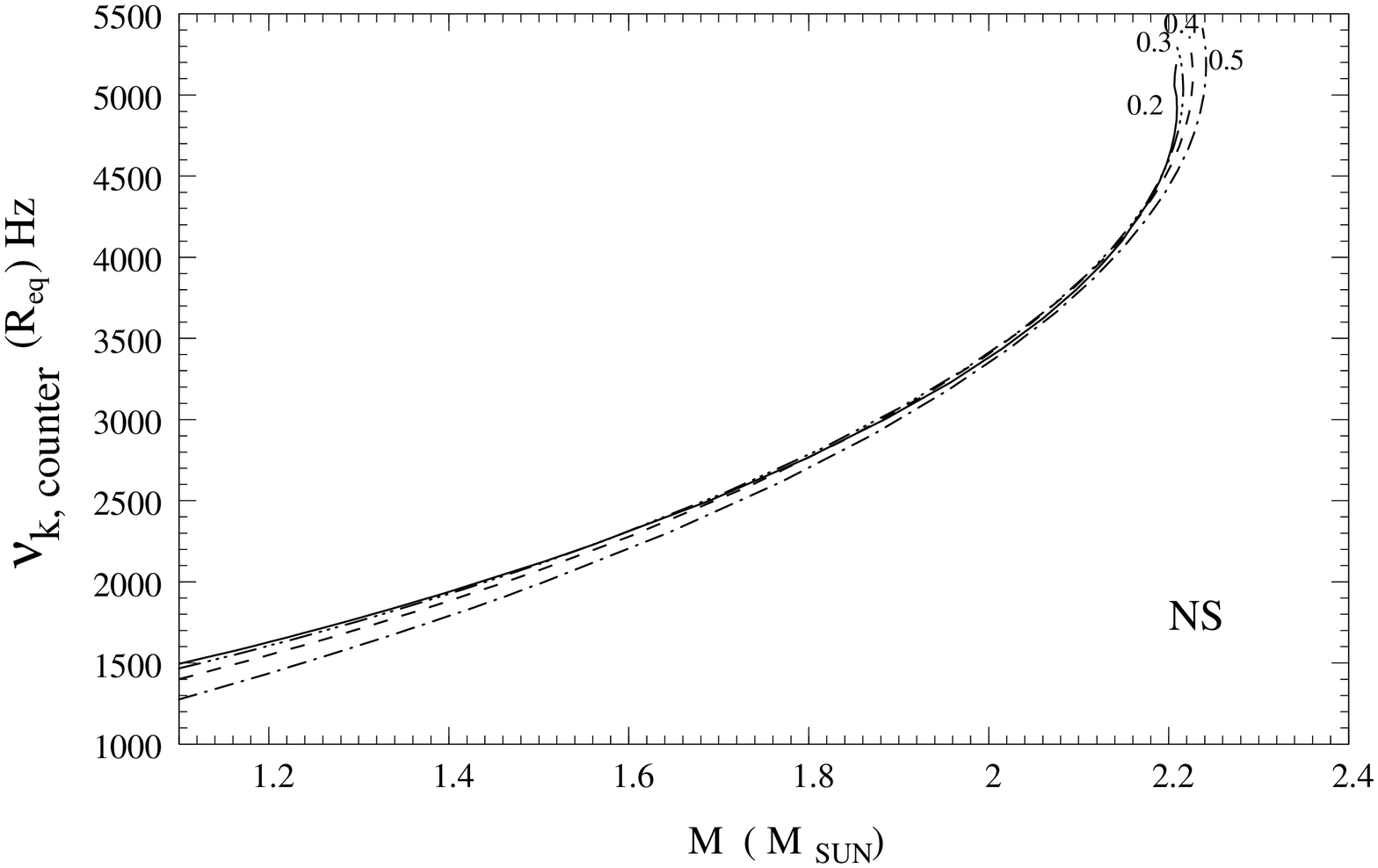,width=8cm}}
\caption{Variation of $\nu_{k,~counter}(R_{eq})$ with the mass for strange stars (upper panel) and neutron stars (lower panel). The parameter is the value of $\Omega$ in units of $ 10^{4}~{\rm sec}^{-1}$ corresponding to $\nu_{spin}$ 318 Hz, 477 Hz, 637 Hz and 796 Hz respectively. The EsoS used are EoS A for strange stars and EoS APR for neutron stars. \label{fig:mnukreqcounter}}
\end{figure}
\begin{figure}
\centerline{\psfig{figure=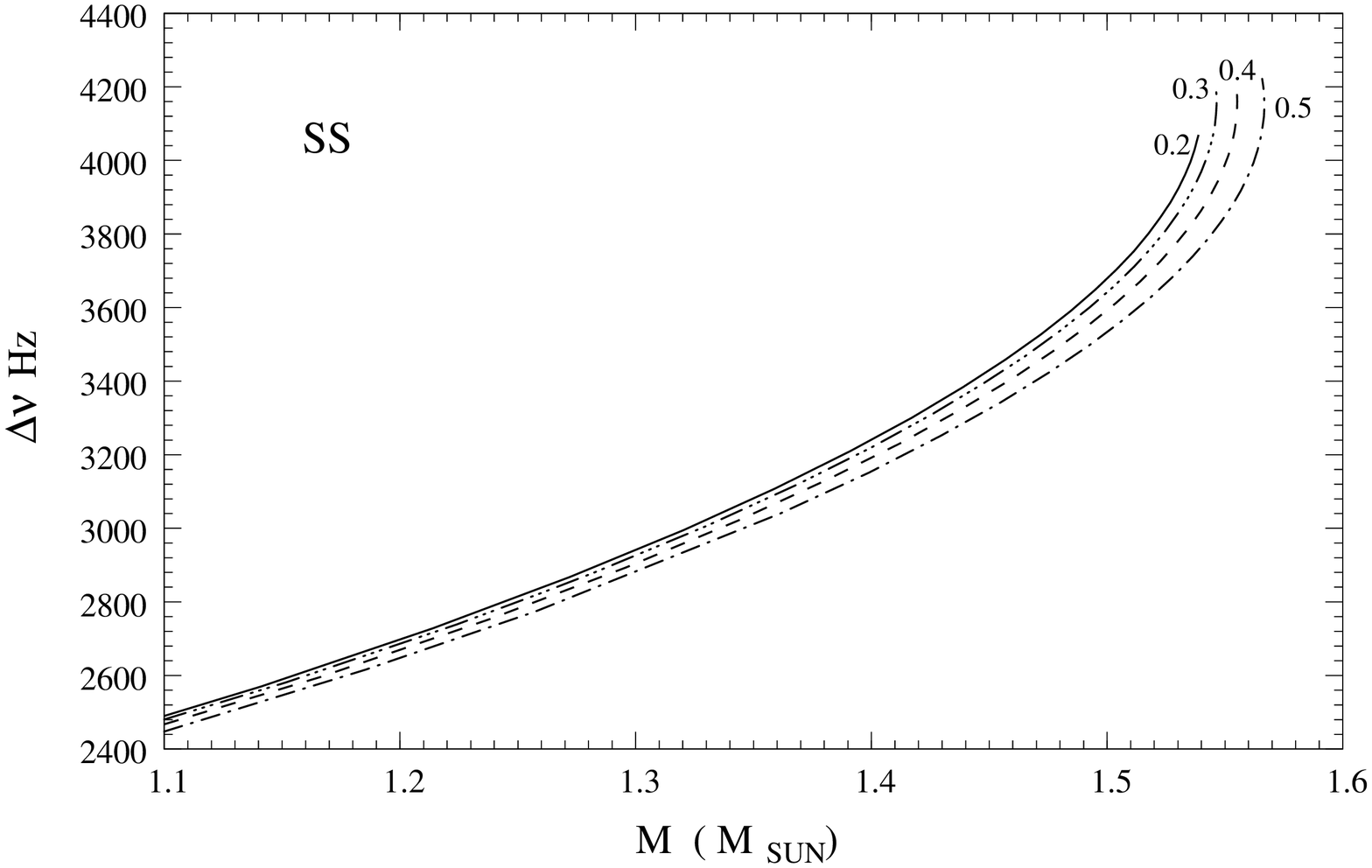,width=8cm}}
\centerline{\psfig{figure=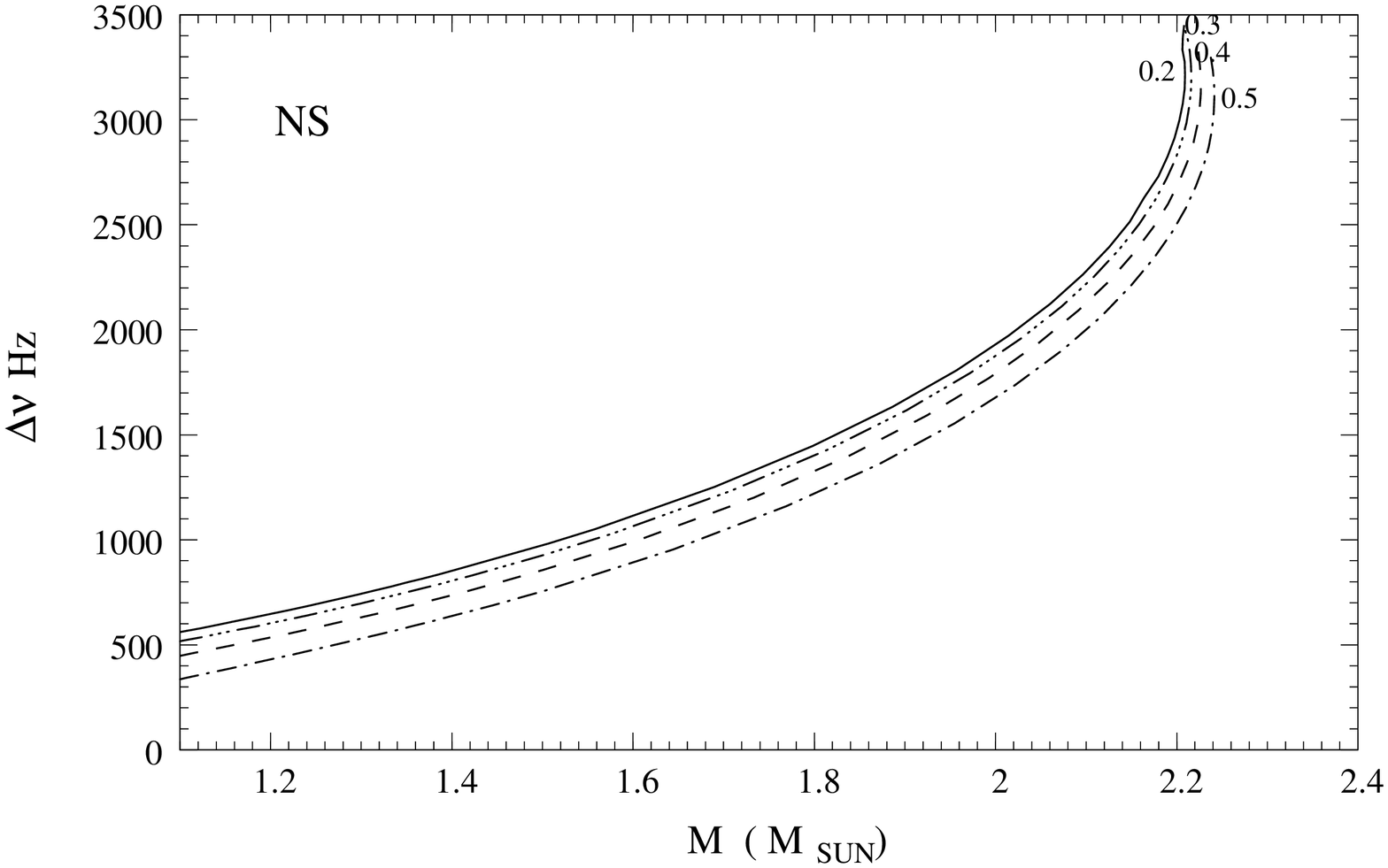,width=8cm}}
\caption{Variation of $\Delta \nu = \nu_{k,~co}(R_{eq}) - \nu_{spin}$ with the mass for strange stars (upper panel) and neutron stars (lower panel). The parameter is the value of $\Omega$ in units of $ 10^{4}~{\rm sec}^{-1}$. The EsoS used are EoS A for strange stars and EoS APR for neutron stars. \label{fig:delnu}}
\end{figure}

In Figs \ref{fig:mnukreqco} and  \ref{fig:mnukreqcounter} I plot $\nu_{k}(R_{eq})$ with the stellar mass for the co-rotating and the counter-rotating motions. $\nu_{k,~co}(R_{eq}) > \nu_{k,~counter}(R_{eq})$ always. Let us concentrate mainly on Fig \ref{fig:mnukreqco} as $\nu_{k,~co}(R_{eq})$ can represent the rotational frequency of both the constituent particles of the star and the accreting material at the equatorial surface of the star whereas $\nu_{k,~counter}(R_{eq})$ can represent only the accreting material. I see that $\nu_{k,~co}(R_{eq})  > \nu_{spin}$ in the chosen range of $M$ and $\nu_{spin}$.  Fig. \ref{fig:delnu} shows that the difference $\Delta \nu~=~\nu_{k,~co}(R_{eq}) - \nu_{spin}$ decreases at higher $\nu_{spin}$ and/or lower $M$. The condition $\Delta \nu~=~0$ is the ``mass shedding limit" as for $\nu_{k,~co}(R_{eq}) < \nu_{spin}$, matter from the star would fly way due to the centrifugal force. The spin frequency where the ``mass-shedding" starts (i.e. $\Delta \nu~ =~0$) is the maximum possible rotational frequency of the star and it is known as the Keplerian frequency of the star ($\nu_{k}^{\ast}= \nu_{spin} = \nu_{k,~co}(R_{eq})$) and the corresponding angular frequency is called as the Keplerian angular frequency ($\Omega_{k}^{\ast}$) of the star. For fixed values of $M$ and $\nu_{spin}$,  ${\Delta \nu}_{SSA} >  {\Delta \nu}_{APR}$. This implies that the ``mass shedding limit" will be reached in case of neutron stars at comparatively a lower value for $\nu_{spin}$ than for strange stars which means that strange stars are more stable than neutron stars against rotations (we have checked that this fact remains true even if one uses BAG model EoS for strange stars). The Keplerian angular frequency $(\Omega_{k}^{\ast})$ for compact stars has been studied in the literature by different groups like Friedman, Ipser \& Parker (1986); Haensel \& Zdunik (1989); Lattimer $et~al.$ 1990 and many other authors. 

Haensel \& Zdunik (1989) proposed an analytic relation between $\Omega_{k}^{\ast}$ and maximum allowable static mass ($M_{max}$) and the corresponding radius ($R_{max}$) 
\begin{equation}
\Omega_{k}^{\ast}~=~7.7 \times 10^{3}\left(\frac{M_{max}}{M_{\odot}} \right)^{0.5} \left(\frac{R_{max}}{10~ \rm{km}}\right)^{-1.5} \label{eq:haen_zdun}
\end{equation}
With their choice of neutron star EsoS, Lattimer $et~al.$ (1990) found  $\Omega_{k}^{\ast}$ to lie between $0.76 \times 10^4 - 1.6 \times 10^4 ~\rm{sec}^{-1}$ from Eq. (\ref{eq:haen_zdun}). Using Bag model EsoS for strange stars, Prakash, Baron \& Prakash (1990) found $\Omega_{k}^{\ast}  \leqslant 1.0 \times 10^4 ~\rm{sec}^{-1}$ for $M > 1.44 M_{\odot}$.

Afterwards, Lattimer and Prakash (2004) gave a more useful expression of $\Omega_{k}^{\ast}$ of a star of mass $M$ and non-rotating radius $R$.

\begin{equation}
\Omega_{k}^{\ast}~=~ 6.6 \times 10^{3}\left(\frac{M}{M_{\odot}} \right)^{0.5} \left(\frac{R}{10~ \rm{km}}\right)^{-1.5} \label{eq:latt_prakash}
\end{equation}

To test these two simple analytical expressions (Eqns. \ref{eq:haen_zdun}, \ref{eq:latt_prakash}), I run the task ``{\it kepler}" in RNS code which produces stellar configurations for stars rotating with $\Omega_{k}^{\ast}$. Using the RNS outputs ($i.e.$, $M$, $R$, $I$, $\Omega_{k}^{\ast}$, I calculate $\nu_{k,~co}(R_{eq})$ from Eqn. \ref{eq:nuk}. At $\Omega_{k}^{\ast}$, I should get $\Omega_{k}^{\ast}~=~2 \pi \nu_{k,~co}(R_{eq})$ (by the definition of $\Omega_{k}^{\ast}$).

\begin{figure}
\centerline{\psfig{figure=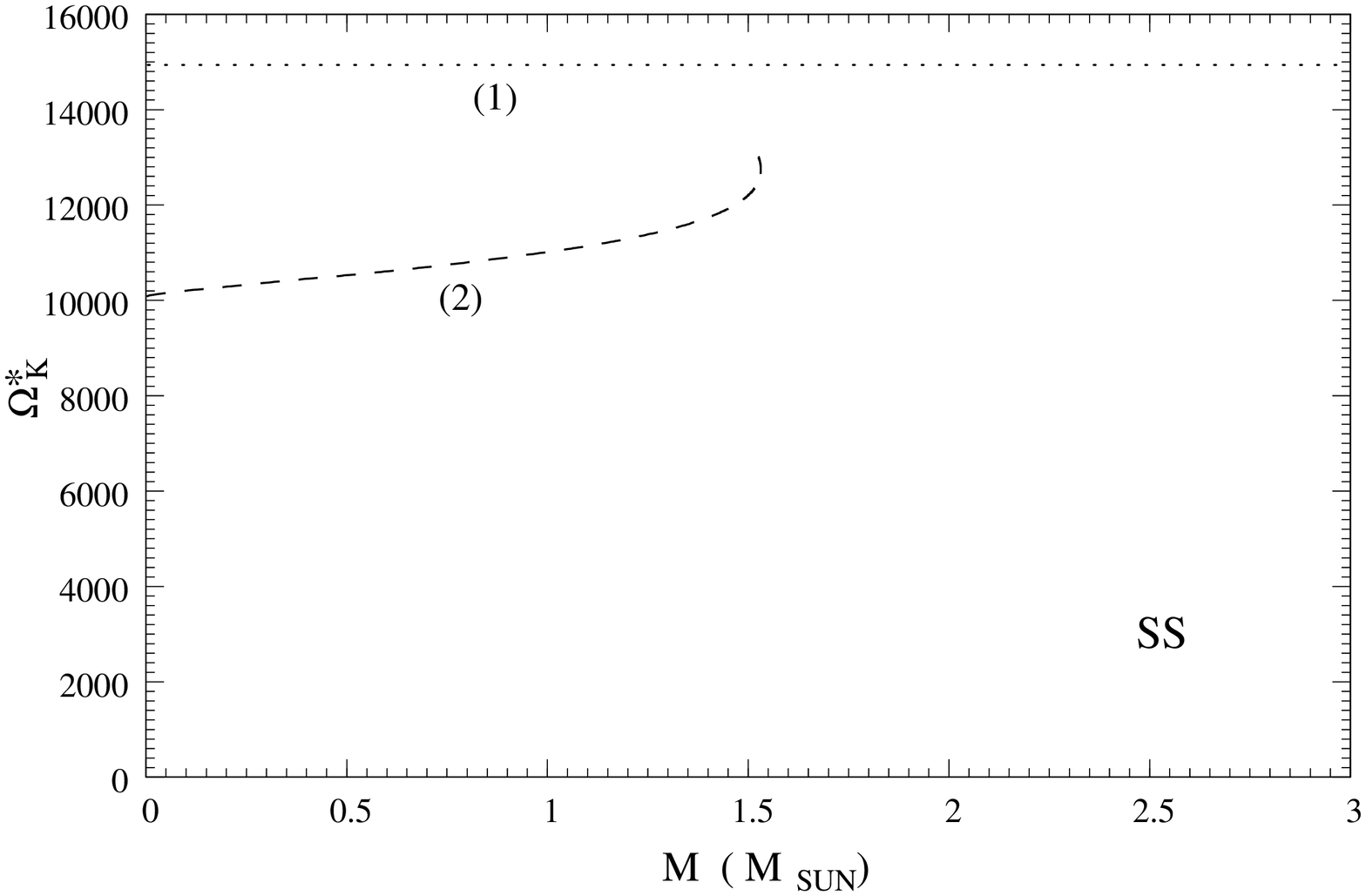,width=8cm}}
\centerline{\psfig{figure=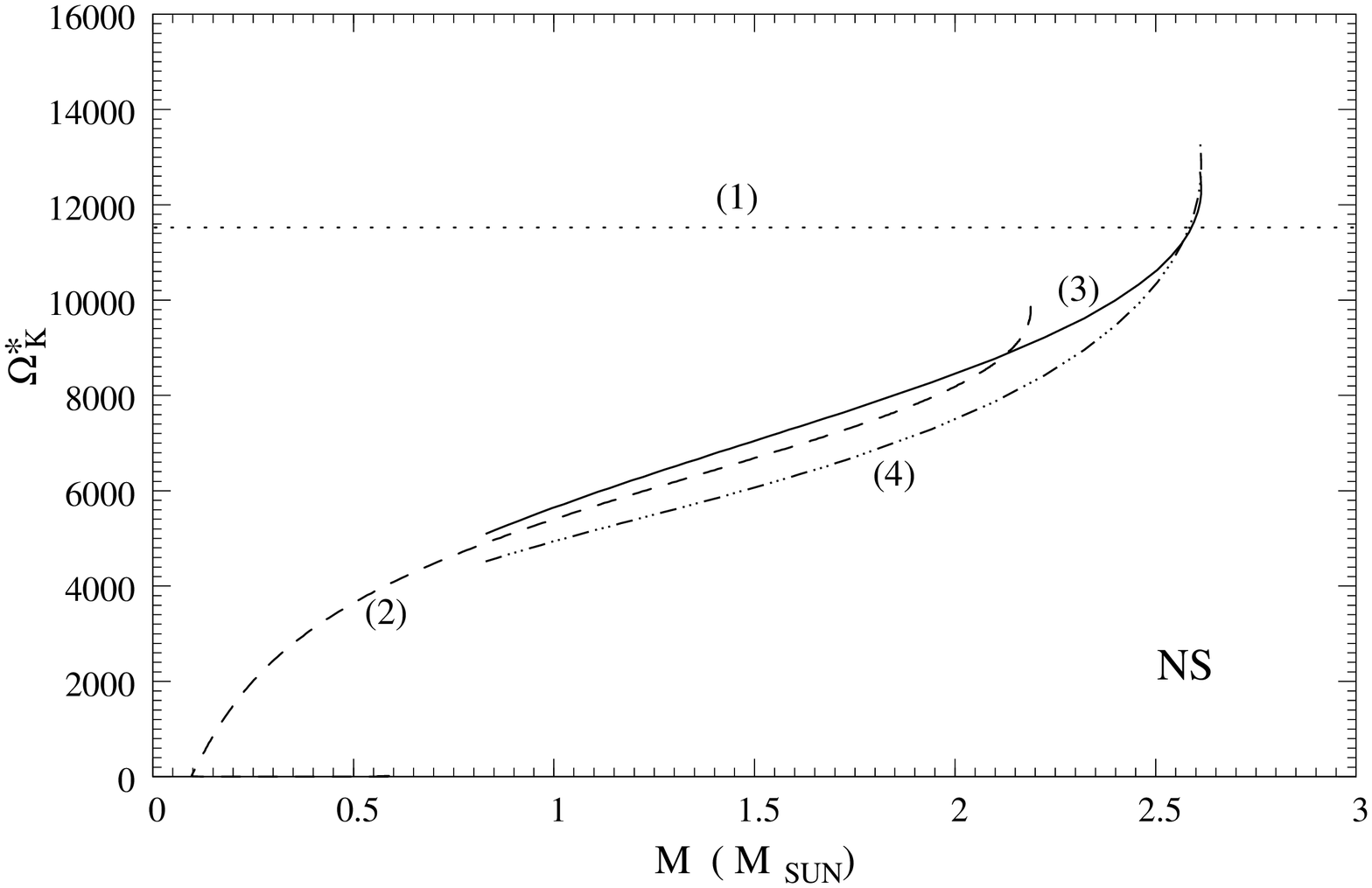,width=8cm}}
\caption{Variation of $\Omega_{k}^{\ast}$ with mass. The line labeled (1) is the maximum limit of $\Omega_{k}^{\ast}$ given by Haensel \& Zdunik (1989). The curve labeled (2) is from the analytical expression given by Lattimer and Prakash (2004). The curve labeled (3) is the output of RNS code and the curve labeled (4) is $2 \pi \nu_{K,~co}(R_{eq})$. The EsoS used are EoS A for strange stars and EoS APR for neutron stars.  \label{fig:omega_kep}}
\end{figure}

In Fig \ref{fig:omega_kep}, I plot the variation of $\Omega_{k}^{\ast}$ with mass. The line labeled (1) is the maximum limit of $\Omega_{k}^{\ast}$ given by Haensel \& Zdunik (1989) $i.e.$ Eqn \ref{eq:haen_zdun}. The curve labeled (2) is from the analytical expression given by Lattimer \& Prakash (2004) $i.e.$ Eqn \ref{eq:latt_prakash}. The curve labeled (3) is the output of the RNS code and the curve labeled (4) is $2 \pi \nu_{k,~co}(R_{eq})$.

For neutron stars (APR EoS) the curves (2), (3), (4) are very close to each other which supports the correctness of both the analytical expression of Lattimer \& Prakash (2004) (Eqn \ref{eq:latt_prakash}) and the Pseudo-Newtonian Potential of Mukhopadhyay \& Misra (2003). The maximum value of $\Omega_{k}^{\ast}$ obtained using Eqn. (\ref{eq:haen_zdun}) is $ \sim 11519~ {\rm sec^{-1}}$. Here the Kerr parameter $a/R_g$ lies around 0.66 which is much greater than the values at lower frequencies (Fig. \ref{fig:ma}) and the neutron stars are very oblate having $R_p/R_{eq} \sim 0.59 - 0.56$. 

For EoS A, the RNS code failed to perform the task ``{\it kepler}" for $\epsilon_c < 1.51 \times 10^{15} {\rm gm~ cm^{-3}}$ and at $\epsilon_c = 1.51 \times 10^{15} {\rm gm~ cm^{-3}}$, I get $M~=~1.61 ~ M_{\odot}$ which is greater than $M_{max}$ for static configuration. Moreover, at this mass, the value of the Kerr parameter $a/R_g$ is greater than 1, which is unphysical. At $\epsilon_c \geqslant 1.58 \times 10^{15} {\rm gm~ cm^{-3}}$ ($M \geqslant 1.72 ~ M_{\odot}$) , $a/R_g$ becomes less than 1, but very high ($\sim 0.9$). So no direct comparison of $\Omega_{k}^{\ast}$ obtained with the RNS code with the analytical expressions are possible and in the figure I only plot the analytical expressions. Using Eqn. (\ref{eq:latt_prakash}), I get $\Omega_{k}^{\ast} = 13086 ~ {\rm sec^{-1}}$ at $\epsilon_c ~=~1.51 \times 10^{15} {\rm gm~ cm^{-3}}$ and increases with the increase of $\epsilon_c$ (or $M$). This value of $\Omega_{k}^{\ast}$ is less than the maximum value of $\Omega_{k}^{\ast}$ obtained using Eqn. (\ref{eq:haen_zdun}) which is $14940~ {\rm sec^{-1}}$. Here the strange stars are very oblate with $R_p/R_{eq} \sim 0.38 - 0.47$. To check whether all these facts are intrinsic to strange star properties or depend upon the particular model of strange stars, I have even checked with the BAG model having model parameters as : $B = 60.0~{\rm MeV / fm^3},~m_s = 150.0~ {\rm MeV},~ m_u = m_d = 0,~\alpha_c = 0.17$ where $B$ is the Bag parameter, $m_s$, $m_u$, $m_d$ are masses of s, u and d quarks respectively giving static $M_{max} = ~1.82~ M_{\odot}$. Here also I get $a/R_g > 1 $ at lower $\epsilon_c$, $a/R_g < 1 $ when $\epsilon_c \geqslant  0.82 \times 10^{15} {\rm gm~ cm^{-3}}$ ($M \geqslant 2.33 M_{\odot}$). Using Eqn. (\ref{eq:latt_prakash}), I get $\Omega_{k}^{\ast} = 8756~ {\rm sec^{-1}}$ at $\epsilon_c ~=0.82~ \times 10^{15} {\rm gm~ cm^{-3}}$ which increases with increase of $\epsilon_c$ (or $M$). This value of $\Omega_{k}^{\ast}$ is less than the maximum value of $\Omega_{k}^{\ast}$ obtained using Eqn. (\ref{eq:haen_zdun}) which is $10695~ {\rm sec^{-1}}$. The strange stars are very oblate having  $R_p/R_{eq} \sim 0.38 - 0.42$.

The probable fastest spin frequency of a neutron star (XTE J1739-285) is  $\nu_{spin}=1122$ Hz or $\Omega = 7049.734 ~ {\rm sec^{-1}}$, which is less than the value of $\Omega_{k}^{\ast}$ of both strange stars and neutron stars as derived from Eqn.( \ref{eq:haen_zdun}) or even less than as derived from Eqn. \ref{eq:latt_prakash} for a canonical value of the stellar mass as $M~=~1.5 ~M_{\odot}$ (see Fig. \ref{fig:omega_kep}). So there is no problem of this star being either a strange star or a neutron star. One needs to conclude about its nature by other observational evidences.

Several other people performed numerical studies on structures of rapidly rotating compact stars like Friedman \& Ipser (1992), Weber \& Glendening (1992); Cook, Shapiro \& Teukolsky (1994); Eriguchi, Hachisu \& Nomoto (1994); Salgado $et~al.$ (1994a,b); Gourgoulhon $et~al.$ (1999); Bhattacharyya, Thampan \& Bombaci (2000); Bombaci, Thampan \& Datta (2000), Gondek-Rosi\'nska $et~al.$ (2000); Bhattacharyya \& Ghosh (2005); Haensel, Zdunik \& Bejger (2008); Haensel $et~al.$ (2009). In these, people usually kept themselves confined in studying the properties of stars like $M$, $\epsilon_c$, $R_{eq}$, $T/W$ etc or at most properties particles rotating at the stellar surface. But the use of the pseudo-Newtonian potential enables us to study the properties of particles orbiting around the star at a much higher distance ($r >> R_{eq}$) and I report quantities like $r_{ms,co}$ and $r_{ms,counter}$, $\nu_{k,co}(r)$, $\nu_{k,counter}(r)$ which are useful to study accretion onto rotating neutron stars or strange stars (see next section). I also report various stellar parameters $M$, $R$, $I$, $R_{p}/R_{eq}$, $a/R_g$, $\nu_{k,co}(R_{eq})$, $\nu_{k,counter}(R_{eq})$ $etc$. There is no previous work where all this parameters were reported together. Also I use one EoS for strange stars and another EoS for neutron stars whereas in the earlier works people discussed either only neutron star rotations or only strange star rotations. I also discuss about kHz QPOs within the scenario of Mukhopadhyay $et~al.$ (2003) in the next section.

\section{Applications in kHz QPO models}
\label{sec:application}

kHz QPOs in the LMXBs are very interesting phenomena. The most popular model to explain kHz QPOs is the beat frequency model (Strohmayer $et~al.$ 1996) which assumes the upper QPO ($\nu_{up}$) as the Keplerian orbital frequency $\nu_{k}(r)$  of the innermost orbit in the accretion disk around the star, the separation between the two peaks $i.e.$ $\Delta \nu_{peak}=\nu_{up}-\nu_{low}$ as $\nu_{spin}$ and the lower QPO  ($\nu_{low}$) as the beat frequency of $\nu_{k}(r)$ and $\nu_{spin}$. This model suggest that for a particular star  $\Delta \nu_{peak}~=~\nu_{spin}$ will remain constant.

However, it has been observed that in many sources, $\Delta \nu_{peak}$ changes with time. This phenomenon discards the reliability of the beat frequency model for the QPOs. Two examples of alternate models can be found in Titarchuk and Osherovich (1999) and Mukhopadhyay et al. (2003). Both of these later two models suggest $\nu_{low}~=~\nu_{k}(r)$. But in the present work, I don't prefer any particular QPO model over the others.  

We take two sample LMXB for which kHz QPOs have been observed, namely KS 1731$-$260 with $\nu_{spin}=524$ Hz and 4U 1636$-$53 $\nu_{spin}=581$ Hz. As none of the above mentioned models for kHz QPOs is established beyond doubt, I fit both $\nu_{up}$ and $\nu_{low}$ to $\nu_{k}(r)$ to find the corresponding radial distance $r=r_{k}$ (from Eq. \ref{eq:nuk}) which I call $r_{k,~up}$ and $r_{k,~low}$ respectively. Mukhopadhyay $et~al.$ (2003) calculated  $r_{k,~low}$ for these two sources taking mass-radius values for non-rotating strange stars and using $a/R_{g}$ ($J$ in their notation) as an parameter. Here I get $r_{k,~up}$ and $r_{k,~low}$ taking exact values of $R_{eq}$ and $a/R_{g}$ as generated by the RNS code with the specific values of $\nu_{spin}$ and chosen values of $M$ (as masses of these two stars are not well determined). QPO frequencies for KS 1731$-$260 has been taken from Wijnands and van der Klis (1997) and those for 4U 1636$-$53 has been taken from Jonker, Mendez \& van der Klis (2002). Note that the QPO frequencies of 4U 1636$-$53 shift with time and hence I have taken only one set of values as an example.

Table \ref{tb:qpo1} shows the rotational parameters for  KS 1731$-$260 and 4U 1636$-$53. Table \ref{tb:qpo2} shows the values of the radius of the Keplerian orbit ($r_k$) obtained by equating $\nu_k (r)$ with QPO frequencies. I choose the stellar mass to be 1.1 $M_{\odot}$, 1.5 $M_{\odot}$ and 1.9 $M_{\odot}$. For a given star ($i.e. ~\nu_{spin}$ is fixed), the values of $r_{ms}$ depend significantly on (a) whether I consider the co-rotating or the counter-rotating motion, (b) on the choice of the EoS and (c) on the chosen value of $M$. But the values of $r_k$ depend significantly only on $M$ when I keep $\nu_{spin}$ and $\nu_k$ fixed.

For KS 1731$-$260, $r_{k,~low}- r_{k,~up} \sim 2$ km for both strange stars and neutron stars (independent of the choice of mass) and for both co-rotating and counter-rotating motions whereas for  4U 1636$-$53 this value is $\sim 4$ km. 

$r_{k,~low}- r_{ms} \sim 4$ to $7$ for KS 1731$-$260 and $\sim 7$ to $10$ km for 4U 1636$-$53 for co-rotating motions, $\sim 2$ to $9$ km for KS 1731$-$260 and $\sim 8$ to $10$ km for 4U 1636$-$53 for counter-rotating motions; whereas $r_{k,~up}- r_{ms} \sim 1$ to $5$ km for KS 1731$-$260 and $\sim 2.5$ to $6$ km for 4U 1636$-$53 for co-rotating motions $r_{k,~up}- r_{ms} \sim -1.5$ to $2.5$ km for KS 1731$-$260 and $\sim -0.5$ to $3.5$ km for 4U 1636$-$53 for counter rotating motions. These values decrease with the increase of $M$ and note that at $M=1.9 M_{\odot}$, $r_{k,~up, ~ counter} < r_{ms}$. Remember, $M_{max} < 1.9 M_{\odot}$ for EoS A for the values of $\Omega$ of KS 1731$-$260 and 4U 1636$-$53.

\begin{table*}
\caption{Rotational parameters for two LMXBs. }
\begin{center}
\begin{tabular}{cc|cccccccc}
\hline \hline
 source  &$\nu_{spin}$ &M & EoS&$R_{eq}$ &$I$&$R_{p}/R_{eq}$& $a/R_g$ &$r_{ms,~co}$ &$r_{ms,~counter}$ \\
   & (Hz)    & ($M_{\odot}$) & & (km) &$(10^{45}~gm~cm^{2})$ &&&(km)&(km) \\
\hline \hline
 KS 1731$-$260  & 524  &1.1 &SSA & 7.344 & 0.564 & 0.982 & 0.174 & 8.801  &  10.656\\
 &   & 1.1 &NS   & 11.874 & 0.975  & 0.918 & 0.301  & 8.081   & 11.297 \\ \hline
4U 1636$-$53 & 581  & 1.1  &SSA & 7.348  & 0.565  & 0.978   & 0.194  & 8.694  & 10.754  \\
 &    & 1.1 &NS   & 12.015  & 0.990 & 0.897 & 0.339  & 7.860  & 11.485 \\
\hline \hline
 KS 1731$-$260  & 524 & 1.5  & SSA &7.688 &0.908 & 0.984 & 0.151 & 12.179 & 14.366 \\
 &   & 1.5 & NS & 11.634 & 1.494 & 0.940 & 0.248 & 11.435 & 15.042 \\ \hline
4U 1636$-$53 & 581  &  1 .5 & SSA & 7.699 & 0.911 & 0.981 & 0.168 & 12.051 & 14.485 \\
 &  & 1.5  & NS & 11.718 & 1.508 & 0.926 & 0.278 & 11.205 & 15.244 \\
\hline \hline
 KS 1731$-$260  & 524 &  &&   &   &  & &  &   \\
 &  &  1.9  &NS &  11.255  & 2.032  & 0.959 & 0.210 & 14.894 & 18.723 \\ \hline
4U 1636$-$53 & 581 &   & &  &  &  &  &  &   \\
 &  & 1.9  & NS & 11.310  & 2.045 & 0.949  & 0.235 & 14.617 & 18.9357 \\
\hline \hline
\end{tabular}
\end{center}
\label{tb:qpo1}
\end{table*}

\begin{table*}
\caption{Keplerian radius ($r_k$) by fitting kHz QPOs for two LMXBs.}
\begin{center}
\begin{tabular}{ccc|cccccc}
\hline \hline
 source &$\nu_{low}$ & $\nu_{up}$ &  $M$ & EoS &$r_{k,~low, ~co}$ &  $r_{k,~low, ~counter}$ &$r_{k,~up, ~co}$ &  $r_{k,~up, ~counter}$\\
     & (Hz) & (Hz) &  ($M_{\odot}$) &  &(km)  & (km) &(km)  & (km)\\
\hline \hline
 KS 1731$-$260   & 898.3 & 1158.6& 1.1 &SSA& 15.135   &  15.341   & 12.923   &  13.245  \\
 &  &    & 1.1 &NS &  15.099  & 15.442   & 12.835   & 13.379   \\ \hline
4U 1636$-$53   & 688 &1013  & 1.1 & SSA & 18.030   & 18.098   & 14.018   & 14.310  \\
 &   & & 1.1 &NS & 18.061 & 18.166 & 13.953 & 14.448 \\
\hline \hline
KS 1731$-$260   & 898.3 & 1158.6 & 1.5 & SSA & 17.071 & 17.477 & 14.758 & 15.255 \\
 &  &    & 1.5  & NS & 16.958 & 17.618 & 14.609 & 15.420 \\ \hline
4U 1636$-$53    & 688 &1013 & 1.5 & SSA & 20.080 & 20.378 & 15.891 & 16.396 \\
 &  & & 1.5  &NS & 20.023 & 20.504 & 15.748 & 16.572 \\  
\hline  \hline
KS 1731$-$260   & 898.3 & 1158.6 &  & &   &    &   &   \\
 &  &    & 1.9 &NS &  18.766  & 19.631   & 16.346  & 17.307  \\ \hline
4U 1636$-$53   &  688 &1013 &   &&   &  &  &   \\
 & &  & 1.9  &NS &  21.896  & 22.662   & 17.507    & 18.530  \\  
\hline \hline
\end{tabular}
\end{center}
\label{tb:qpo2}
\end{table*}

\section{Discussion}
\label{sec:conclusion}

We study the rotational parameters for strange stars and neutron stars using the RNS code which consider the generalized axisymmetric metric for rotating stars. I find that the rotational parameters like $M_{max}$, $R_{eq}$, $I$, $R_{p}/R_{eq}$, $a/R_{g}$ depend on $\nu_{spin}$ and the choice of the EoS. The value of $r_{ms}$ depends on the stellar mass, the choice of the EoS and whether the motion is co-rotating or counter-rotating. The dependence of $\nu_{k}(r)$ on the choice of EoS, $\Omega$ and the direction of motion is prominent only at low $r$ ($ \leqslant R_{eq}$). That is why, the values of $r_k$ as obtained by fitting the kHz QPO frequencies do not depend much on the choice of the EoS and on the direction of motion as here $r_k > R_{eq}$.

The beat frequency model suggests that $r_{k,~up}$ is the radius of the innermost Keplerian orbit of the ($r_{in}$) of the accretion disk whereas the models by Titarchuk and Osherovich (1999) and Mukhopadhyay $et~al.$ (2003) suggest that $r_{k,~low}=r_{in}$. The disk parameter $r_{in}$ can be estimated by X-ray spectral analysis. As relativistic broadening is more dominant at the innermost edge of the disk, the existence of a relativistically broadened iron K$\alpha$ line helps one to determine the value of $r_{in}$ (Reis, Fabian \& Young 2009; Di Salvo $et~al.$ 2009 and references therein). It is clear from the table \ref{tb:qpo2} that for a fixed set of $M$ and $\nu_{spin}$, $r_{in}$ is different for different EsoS, but the difference is always $<1\%$, whereas presently $r_{in}$ is measured only accuracy up to $20 \%$ (Reis, Fabian \& Young 2009; Di Salvo $et~al.$ 2009). Moreover, one should remember that this determination of $r_{in}$ depends upon the QPO model which is not beyond doubt at the present moment. Indeed there is a strong need of better explanation of kHz QPOs to understand various features like the shift of the peaks, their correlations (Belloni, M\'endez and Homan 2007; Yin $et~al.$ 2007), side-bands (Jonker, M\'endez and van der Klis 2005) etc. Considering all these facts, I conclude that, to constrain dense matter Eos by measuring $r_{in}$ using X-ray spectral analysis, I need much better accuracy and hopefully future advanced technology like ASTROSAT will provide us such ultra-high accuracy. 

In this work I have primarily confined myself to a maximum value of $\nu_{spin}$ as 796 Hz. But then I have also studied the properties of the compact stars if they rotate with the maximum spin frequency $i.e.$ the Keplerian spin frequency.


\begin{thebibliography}{99}

\bibitem{brdd} Bagchi, M., Ray, S., Dey, J., \& Dey, M., 2006,  Astron. \& Astrophys., 450, 431.
\bibitem{apr} Akmal, A., Pandharipande, V. R., Ravenhall, D. G., 1998, Phys. Rev. C 58, 1804.
\bibitem{bagchidns} Bagchi, M., Dey, J., Konar, S., Bhattacharya, G., Dey, M., arXiv:astro-ph/0610448.
\bibitem{bardeen72} Bardeen, J., Press, W. H., Teukolsky, S. A., 1972, Astrophys. J. 178, 347.
\bibitem{belloniqpo} Belloni, T., Mendez M., Homan J., 2007, Mon. Not. Roy. Astron. Soc. 376, 1133.
\bibitem{bhat_ghosh}Bhattacharyya, A., Ghosh, S., 2005, arXiv:astro-ph/0506202.
\bibitem{sudip2000} Bhattacharyya, S.,  Thampan, A. V., Bombaci, I., 2001, Astron. \& Astrophys 372, 925.
\bibitem{bom_2000} Bombaci, I., Thampan, A. V., Datta, B., 2000, Astrophys. J. 541, L71.
\bibitem{cook94} Cook, G. B., Shapiro, S. L., Teukolsky, S. A., 1994, ApJ, 424, 823.
\bibitem{salvo09} Di Salvo, T., D' Ai, A., Burderi, L., $et~al.$ 2009, arXIv:0904.3318.
\bibitem{erig94} Eriguchi, Y., Hachisu, I., Nomoto, K., 1994, MNRAS, 266, 179.
\bibitem{fried_ips_park} Friedman, J. L., Ipser, J. R, Parker, L., 1986, Astrophys. J. 304, 115.
\bibitem{fried_ips92}Friedman J. L., Ipser, J. R, 1992, Phil. Trans. R. Soc. Lond. A, 340, 391.
\bibitem{dorota2000}Gondek-Rosi\'nska, D., Bulik, T., Zdunik, L., Gourgoulhon, E., Ray, S., Dey, J., Dey, M., 2000, Astron. \& Astrophys. 363, 1005.
\bibitem{gour99} Gourgoulhon, E., Haensel, P., Livine, R., $et~al.$, 1999, A \& A, 349, 851.
\bibitem{haensel89} Haensel, P., Zdunik, J. L., 1989, Nature 340, 617.
\bibitem{haensel08}Haensel, P., Zdunik, J. L., Bejger  M., 2008, New Astron. Rev. 51, 785.
\bibitem{haensel09}Haensel, P., Zdunik, J. L., Bejger, M., Lattimer, J. M., 2009, arXiv:0901.1268
\bibitem{hessels06} Hessels, J. W. T., Ransom, S. M., Stairs, I. H., 2006, Science, 311, 1901. 
\bibitem{jmk} Jonker, P. G., M\'endez, M., van der Klis, M., 2002, Mon. Not. Roy. Astron. Soc 336, 1.
\bibitem{kaaret} Kaaret, P., in 't Zand, J. J. M., Brabdt, S. $et~al.$, 2007, Astrophys. J. 657, L97.
\bibitem{kom} Komatsu, H., Eriguchi, Y., Hachisu, I., 1989, Mon. Not. Roy. Astron. Soc. 237, 355.
\bibitem{klahn}Klahn, T., Blaschke, D., Typel, S., $et~al.$, 2006, Phys.Rev. C 74, 035802.
\bibitem{lm04} Lattimer, J. M., Prakash, M. , 2004, Science 304, 536.
\bibitem{lm07} Lattimer, J. M., Prakash, M. , 2007, Physics Reports 442, 109.
\bibitem{lpmy} Lattimer, J. M., Prakash, M. , Masak, D., Yahil, A., 1990, Astrophys. J. 355, 241.
\bibitem{ls} Lattimer, J. M., Schutz, B. F. , 2005, Astrophys. J. 629, 979.
\bibitem{liqpo}Li, X. D., Ray, S., Dey, J., Dey, M., Bombaci, I., 1999, Astrophys. J. 527, L51.
\bibitem{banipot} Mukhopadhyay, B., Misra, R., 2003, Astrophys. J. 582, 347.
\bibitem{baniqpo} Mukhopadhyay, B., Ray, S., Dey, J., Dey, M., 2003, Astrophys. J. 584, L83.
\bibitem{ozel} \"Ozel, F., 2006, Nature, 441, 1115.
\bibitem{ozel} \"Ozel, F., arXiv:0810.1521.
\bibitem{reis09}Reis, R. C., Fabian, A. C., Young, A. J., 2009, arXiv:0904.2747.
\bibitem{salgado94a} Salgado, M., Bonazzola, S., Gourgoulhon, E., Haensel, P., 1994a, A \& A, 291, 155. 
\bibitem{salgado94b} Salgado, M., Bonazzola, S., Gourgoulhon, E., Haensel, P., 1994b, A \& A Suppl., 108, 455.
\bibitem{strohqpo}
Strohmayer, T. E., Zhang, W., Swank, J. H., $et~al.$ 1996, Astrophys. J. 469, L9.
\bibitem{titqpo} Titarchuk, L., and Osherovich, V., 1999, Astrophys. J. 518, L95.
\bibitem{web_glend_1992} Weber, F., Glendenning, N., 1992, Astrophys. J. 390, 541.
\bibitem{rdyklis} Wijnands, R. A. D. and van der Klis, M., 1997, Astrophys. J. 482, L65.
\bibitem{yinqpo} Yin, H. X., Zhang, C. M., Zhao, Y. H., $et~al.$ 2007, Astron. and Astrophys. 471, 381.
\bibitem{zhang07}Zhang, C. M., Yin, H. X., Kojima, Y., $et~al.$, 2007, Mon. Not. Roy. Astron. Soc. 374, 232.

\end{thebibliography}
\end{document}